\documentclass[fleqn,usenatbib]{mnras}

\usepackage[T1]{fontenc}
\usepackage{ae,aecompl}

\usepackage{graphicx}	
\usepackage{amsmath,comment}	
\usepackage{amssymb}	
\usepackage{xcolor}
\usepackage{soul}
\setstcolor{red}
\setul{}{2pt}

\def\galaxia{{\sc galaxia}}
\def\surfs{{\sc surfs}}

\newcommand{\kms}{\ifmmode  \,\rm km\,s^{-1} \else $\,\rm km\,s^{-1}  $ \fi }
\newcommand{\kpc}{\ifmmode  {\rm kpc}  \else ${\rm  kpc}$ \fi  }  
\newcommand{\Msun}{\ifmmode {\rm M_{\odot}} \else ${\rm M_{\odot}}\odot\odot\odot\odot$ \fi} 

\newcommand{\rsigma}{\sigma_r}
\newcommand{\thsigma}{\sigma_\theta}
\newcommand{\phsigma}{\sigma_\phi}

\newcommand{\LCDM}{$\Lambda{\rm CDM}$}

\defcitealias{2005ApJ...635..931B}{BJ05}
\defcitealias{2008ApJ...689..936J}{J08}

\title[Mass reconstruction of the MW analogues]{Jeans that fit: weighing the mass of the Milky Way analogues in the \LCDM\ universe}
\author[P. R. Kafle et al.]
{Prajwal R. Kafle,$^{1}$ \thanks{E-mail: pkafauthor@gmail.com, prajwal.kafle@uwa.edu.au}
 Sanjib Sharma,$^{2}$ 
 Aaron S. G. Robotham,$^{1}$ 
 Pascal J. Elahi, $^{1}$ 
 \newauthor
 and 
 Simon P. Driver,$^{1}$
\newauthor
\\
$^{1}$ ICRAR, The University of Western Australia,35 Stirling Highway, Crawley, WA 6009, Australia\\
$^{2}$ SIfA, A28 School of Physics, The University of Sydney, Sydney, NSW 2006, Australia\\
}

\begin{document}
\label{firstpage}
\pagerange{\pageref{firstpage}--\pageref{lastpage}}
\maketitle

\begin{abstract}
The spherical Jeans equation is a widely used tool for dynamical study of gravitating systems in astronomy. 
Here we test its efficacy in robustly weighing the mass of Milky Way analogues, given they need not be in equilibrium or even spherical. 
Utilizing Milky Way stellar halos simulated in accordance with \LCDM\ cosmology by Bullock and Johnston (2005) and analysing them under the Jeans formalism, 
we recover the underlying mass distribution of the parent galaxy, within distance $r/\kpc\in[10,100]$, with a bias of $\sim12\%$ and a dispersion of $\sim14\%$. 
Additionally, the mass profiles of triaxial dark matter halos taken from the \surfs\ simulation, within scaled radius $0.2<r/r_\text{max}<3$, are measured with a bias of $\sim-2.4\%$ and a dispersion of $\sim10\%$. 
The obtained dispersion is not because of Poisson noise due to small particle numbers as it is twice the later.  
We interpret the dispersion to be due to the inherent nature of the \LCDM\ halos, for example being
aspherical and out-of-equilibrium. 
Hence the dispersion obtained for stellar halos sets a limit of about $12\%$ (after adjusting for random uncertainty) on the accuracy with which the mass profiles of the Milky Way-like galaxies can be reconstructed using the spherical Jeans equation. This limit is independent of the quantity and quality of the observational data.  
The reason for a non zero bias is not clear, hence its interpretation is not obvious at this stage. 
\end{abstract}

\begin{keywords}
Galaxy:kinematics and dynamics -- Galaxy:halo
\end{keywords}

\section{Introduction}
A dynamical model is an essential tool for robust measurement of a mass of a galaxy.
In particular, for the Milky Way (Galaxy) there are a variety of techniques to measure its mass, e.g., the orbital evolution of satellite galaxies \citep{1982MNRAS.198..707L,2007ApJ...668..949B}; the timing argument \citep{2008MNRAS.384.1459L}; the escape velocity \citep{2007MNRAS.379..755S,2014A&A...562A..91P}; directly fitting tracer kinematics and spatial properties with some parametric models \citep{1998MNRAS.294..429D,2011MNRAS.414.2446M,2013A&A...549A.137I} 
or distribution function \citep{1999MNRAS.310..645W,2015MNRAS.453..377W,2015MNRAS.454.3653B,2016ApJ...829..108E}; or calibrating from numerical simulations \citep{2008ApJ...684.1143X,2013ApJ...773L..32R}.
For a more comprehensive list of the literature that attempts to measure the dynamical mass of the MW galaxy see review articles \cite{2014RvMP...86...47C} and \cite{2016ARA&A..54..529B}.

All these mass modelling schemes are known to have their own shortcomings and a common conclusion that can be inferred from the literature is that the 
typical precision in the mass measurement of the Galaxy is roughly $30 \%$ at best \citep[][see Fig. 1]{2015MNRAS.453..377W}.
Generally, the systematics or biases introduced by the method of choice are largely ignored or inaccessible,
and a detailed comparison of the different methods and their relative performances is the matter of a separate review.
Inherently, the main challenge in our ability to achieve higher accuracy in these mass estimates is due to the lack of the tangential information of the halo tracers.
With the ESA's {\it Gaia} mission \citep{2016A&A...595A...2G} we have now entered an era of precision astrometry.
The upcoming DR2 and the subsequent data releases of the mission is expected to deliver 
the full phase-space information of the vast number of stars  \citep{2002Ap&SS.280....1P,2005MNRAS.359.1287B}
that is anticipated to allow us to measure the dynamical properties of the Galaxy, 
especially its mass with unprecedented accuracy.

But, the {\it Gaia} data may not be a panacea as under the current framework of hierarchical structure formation, spiral galaxies like the Galaxy are considered to be formed through the merger of multiple proto-systems. 
In the Galaxy, there are both qualitative and quantitative observational evidence for the presence of a plethora of unrelaxed dynamical structures such as stellar streams and perished dwarf-galaxies \citep[e.g.][]{1993ARA&A..31..575M,2006ApJ...642L.137B,2008ApJ...680..295B,2008A&ARv..15..145H,2009ApJ...698..567S,2011MNRAS.417.2206C}, which supports the scenario of hierarchical galaxy formation. 
Unrelaxed substructures can violate the dynamical consistency of models as well as assumptions of azimuthal symmetry and sphericity. 
Such features perturb the matter distribution of the host galaxy causing the gravitational field to vary with time, which further complicates the mass measurement. 
While fitting the galaxy whether one should include or mask these sub-structures is still open to discussion. 
Moreover, the {\it Gaia} data also comes with uncertainties, which become even more significant at larger distances where the stellar halo is known to dominate.
For example, even for RR Lyrae stars at galacto-centric radius of $40\,\kpc$, the percentage error in {\it Gaia} distance measurement is $>100\%$ and error in the tangential velocity is of the order of at-least $50\kms$ \citep{2005MNRAS.359.1287B,2013ApJ...778L..12P}.

An another approach for measuring the mass of the galaxy is a moment-based method using the \cite{1915MNRAS..76...70J} equation, which we focus on this paper.
If the first and second order moments of the velocity of the mass tracers and their stellar density distribution, which are usually observables, are known, we can use the Jeans equation to infer the gradient of the underlying gravitational potential and hence, the mass of the system.
In astrophysics, the Jeans formalism has often been used for the dynamical study of gravitational systems, ranging from the galaxy superclusters \citep{1999ApJ...512L...9M}, clusters \citep{1997ApJ...476L...7C, 1997ApJ...485L..13C,2003ApJ...585..205B}, groups \citep{2007MNRAS.375..313F,2015MNRAS.453.3848D}, distant galaxies \citep{1983ApJ...266...58T,1990ApJ...361...78B,2008MNRAS.390...71C} 
to stellar clusters \citep{1995AJ....109..209G,2001ApJ...559..828C,2010MNRAS.406.1220W,2014MNRAS.437.3172D} and satellite galaxies \citep{2007ApJ...663..948G,2009MNRAS.394L.102L,2009ApJ...704.1274W,2013NewAR..57...52B,2017MNRAS.470.2034D}. 
It has also been extensively used for the Galaxy \citep{2005MNRAS.364..433B,2010ApJ...720L.108G,2011A&A...531A..82S,2012ApJ...761...98K,2014ApJ...794...59K} and also recently in the neighbouring M31 galaxy \citep{2016MNRAS.460.2720K}.
Generally, a key assumption made in these analysis is that the system is in dynamical equilibrium, without this assumption no information on the underlying gravitational potential can be obtained.
Furthermore, \cite{2013NewAR..57...29B} argues that our strategy must be to start from the assumption of a steady state and then to use perturbation theory to understand how time-dependent effects modify a steady-state model.
However, a crucial question to ask is to what extent do the unrelaxed substructures present in galaxies upset the estimated mass of the host using the Jeans analysis?
This is the main question we aim to find answer to in this paper.

More recently there have been attempts to test the efficacy of different methods in robustly measuring the mass distributions of the Milky Way (MW) like galaxies using spatial and kinematic data. 
Such as \cite{2016MNRAS.456.3456C} demonstrate the ability of cylindrical Jeans equation in reconstructing the surface density, the vertical force, and other disk parameters. 
Similarly, \cite{2015MNRAS.453..377W} test the accuracy of a phase-space distribution function in measuring the mass of the MW dark matter halo.
Besides, \cite{2015ApJ...801...98S} investigate the application of action-space clustering of tidal streams, \cite{2013MNRAS.433.1826S} test stream modelling algorithm in action-angle space, 
and  \cite{2012MNRAS.420.2562A} demonstrate the application of tracer mass formalism in correctly recovering the underlying potential of galaxies. 
Similarly, in this paper we test the \cite{1915MNRAS..76...70J} formalism, in particular of its spherical form in reconstructing the mass distribution of the N-body models of the Milky-Way.

\begin{figure}
  \centering
    \includegraphics[width=1\columnwidth]{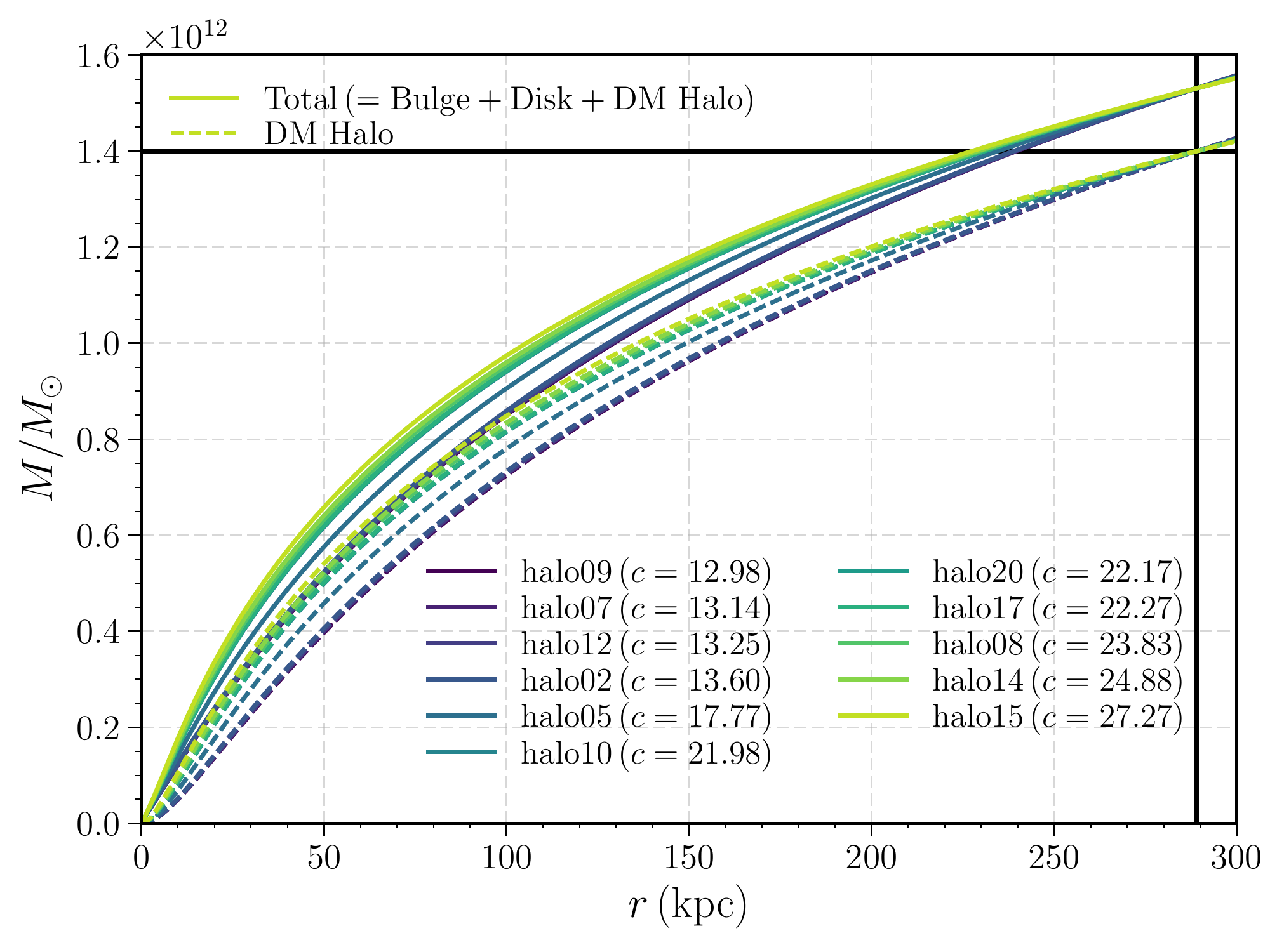}
      \caption{Mass profiles of the 11 \protect\citetalias{2005ApJ...635..931B} \LCDM\ halos and the parent galaxy. Dashed lines show the DM halo mass whereas the solid lines show the combined mass due to bulge, disk and halo. Vertical and horizontal black solid lines correspond to the virial properties of the halos for the concentration parameter of $c=14$. The systematic drift in the runs is due to the concentration $c$ varying among the halos.}
\label{fig:bnjallhalomass}
\end{figure}
With the advent of large spectroscopic surveys such as Sloan Extension for Galactic Understanding and Exploration \citep[SEGUE,][]{2009AJ....137.4377Y}, 
LAMOST Experiment for Galactic Understanding and Exploration \citep[LEGUE,][]{2012RAA....12..735D} and
GALactic Archaeology with HERMES \citep[GALAH,][]{2015MNRAS.449.2604D,2017MNRAS.465.3203M}
we now have a large collection of stars of different stellar types tracing snapshots of the different age populations within the Galactic stellar halo. 
\cite{2005MNRAS.364..433B,2014ApJ...794...59K} have combined different stellar populations such as field horizontal branch stars, giant stars etc to measure the mass of the Galaxy. 
If these different sub-populations of stars trace similar spatio-kinematics loci, or in other words they have been orbiting in the galaxy for 
sufficient time to become fully mixed, it is sensible to combine and treat them as a single population from a dynamical point of view.
Hence, it is also instructive to understand possible bias in the mass measurement resulting from the use of different stellar populations as a tracer.

The rest of this paper is organized as follows. 
Section~\ref{sec:data} describes the data sets obtained from \cite[][hereafter BJ05]{2005ApJ...635..931B} and \cite[][hereafter J08]{2008ApJ...689..936J} simulations used in the paper. 
Section~\ref{sec:jeans} provides the formulary for the Jeans formalism and also, a test case is given in Section~\ref{sec:hqtest}.
In Section~\ref{sec:results} we present our results and discuss them.
Finally, we summarize our findings in Section~\ref{sec:conclusions}. 

\section{Data}\label{sec:data}
\begin{figure*}
  \centering 
  \includegraphics[width=1.9\columnwidth]{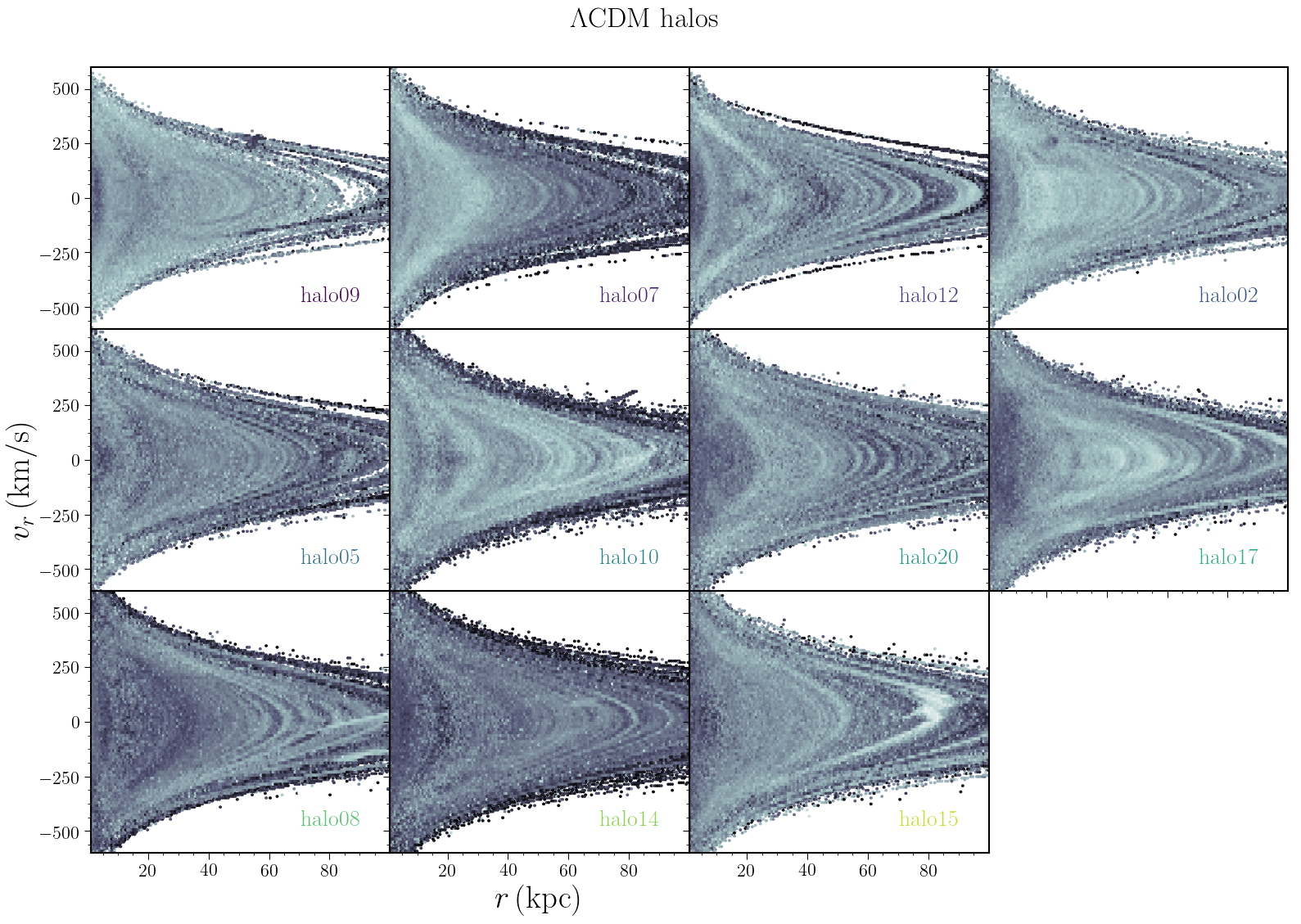}
  \includegraphics[width=1.9\columnwidth]{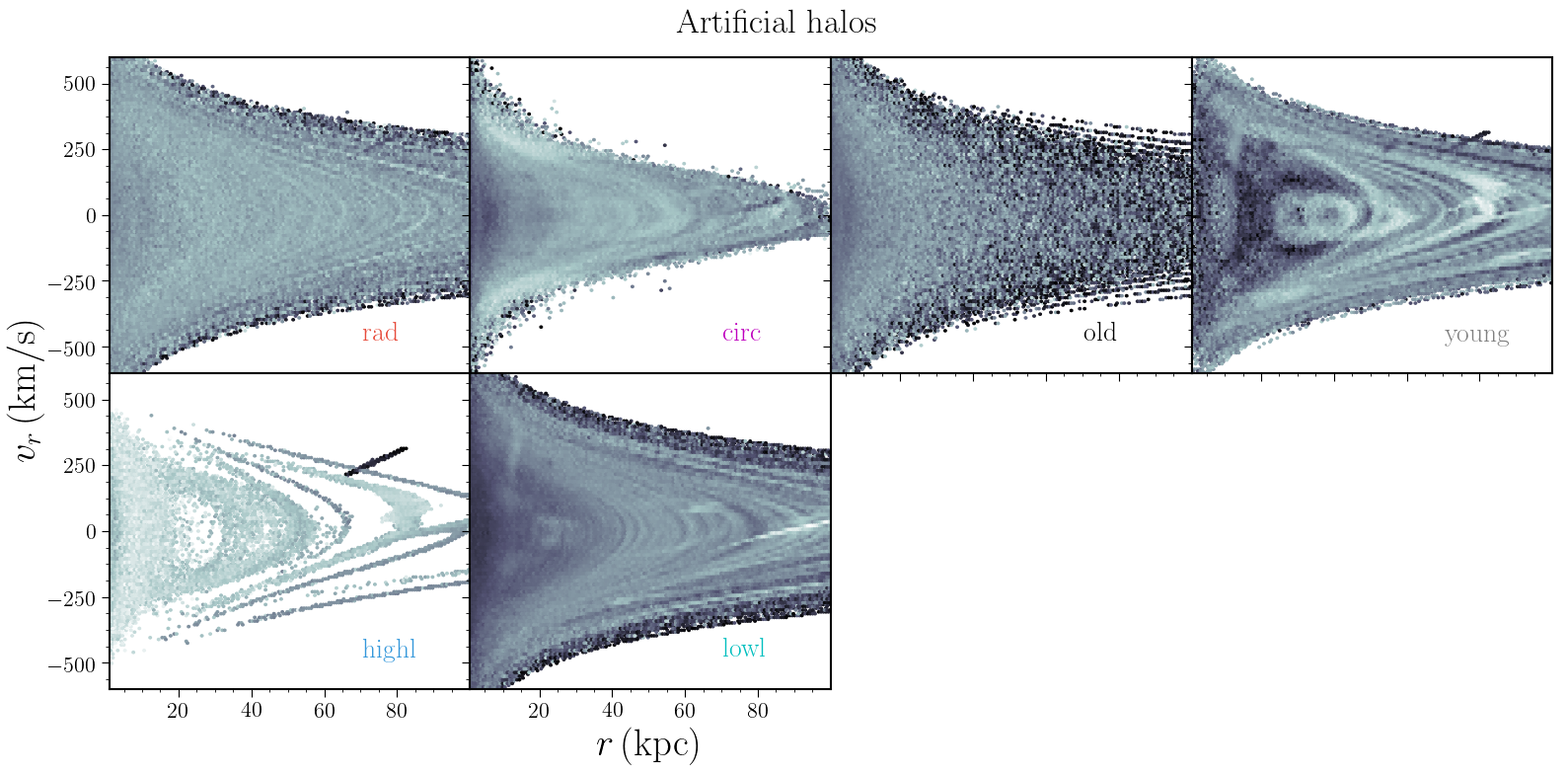}
      \caption{Radial phase-space diagrams ($v_r$ versus halo-centric radius $r$) of all the simulated stellar halos from the \protect\citetalias{2005ApJ...635..931B} (top panels) and \protect\citetalias{2008ApJ...689..936J} (bottom panels) simulations within $r/\kpc\in[1,100]$. }
             \label{fig:rvr}
\end{figure*}
Over the last decade there has been substantial development in simulating stellar halos in a cosmological context.
Full hydrodynamical simulations of galaxies including star formation and feedback recipes have been done 
\citep{2006MNRAS.365..747A,2008MNRAS.391.1685S,2009ApJ...702.1058Z,2014MNRAS.444.1518V,2016MNRAS.457..844F,2016MNRAS.457.1931S}.
However, modelling the stellar halo of galaxies is still a challenging task as it is intrinsically faint \citep{2002ARA&A..40..487F,2008A&ARv..15..145H,2009Natur.461...66M}. 
The highest resolution simulations only resolve stellar masses of $10^4-10^5 M_\odot$, 
and are unable to discern the features in the stellar halo. 
Pushing this further to resolve the faint structures within the stellar halo would require enormous computation power. 
Such simulations will be further complicated by the gas inflow into galaxies, the star-formation physics and its feedback, effects that are poorly understood.
Alternatively, one could utilise models from hybrid techniques \citep{2005ApJ...635..931B,2008MNRAS.391...14D,2010MNRAS.406..744C} 
that use collision-less simulations to track the evolution of dynamical tracers in an analytical potential and also, incorporate star formation processes in a semi-analytical form.
In this paper we use the \citetalias{2005ApJ...635..931B} and  \citetalias{2008ApJ...689..936J} suites of simulations.
These are readily available and have been widely used in the MW studies for its ability to reproduce substructures in great detail \citep{2008ApJ...680..295B,2009ApJ...703.1061S,2011ApJ...728..106S}. 

The \citetalias{2005ApJ...635..931B} and \citetalias{2008ApJ...689..936J} simulations adopt a two-phase approach:
a simulation phase where the N-body simulations of the dark matter (in a short form, DM) satellites 
with known binding energy, accretion time, and eccentricity adopted from the \LCDM\ models are run 
in an analytic disk, bulge and halo potential representing the parent galaxy; 
and a prescription phase where star particles are embedded within the cores of the accreted satellite dark halos.
In other words, the satellite stellar distribution is not modelled during the N-body simulations, 
rather to each DM particle within a satellite the stars are painted on subsequently.
Semi-analytical prescriptions are used to assign a star formation history to each accreted satellite
and the luminosity function of the satellite galaxies is assumed to follow King model.
Afterwards, the orbit of the accreted satellites that contain a significant stellar component are tracked from entry into the parent halo up to the present time.
Only satellite halos more massive than $10^8 M_\odot$ are tracked as the smaller systems never contain enough number of star particles and 
thus do not contribute significantly to the build up of the tracer population of the parent halo.
Time since accretion, luminosity and the eccentricity of the orbit of a satellite are the three key properties that define the course of its accretion.
A full accretion history of the tracer population is a result of the combination of these three properties of all the contributing satellites.

Finally, the above hybrid simulations provide us with two types of dynamical tracers, namely (a) accreted DM particles and (b) stellar particles (essentially accreted DM particles with luminosity weights), originating from dissolving satellites.
Note, the density distribution of the stellar particles shows more substructures than the DM particles as the stellar particles are more tightly bound to the accreting satellites. 
Since, the stellar particles are the ones that have the observational relevance, for our analysis we primarily make use of them. 
For the sake of completeness the results with DM tracer particles are presented in the Appendix~\ref{sec:dmcase}.

\subsection{Simulated stellar halos with \LCDM\ accretion history}
The \citetalias{2005ApJ...635..931B} simulation comprise a suite of 11 random realizations of the MW stellar halo, namely,
\begin{itemize} 
\renewcommand{\labelitemi}{\scriptsize$\blacksquare$} 
\item \emph{halos}: 02, 05, 07, 08, 09, 10, 12, 14, 15, 17 and 20\footnote{we keep the numbering convention of the halos  
            identical to \url{http://galaxia.sourceforge.net/Galaxia3pub.html} \citep[{\galaxia},][]{2011ApJ...730....3S}, 
            which are originally labelled as halos 1-11 in \citetalias{2005ApJ...635..931B}.},
\end{itemize}
here onward we collectively refer to them as \LCDM\ halos.

As we mentioned earlier the potential of the parent galaxy is comprised of three components ---
\cite{1990ApJ...356..359H} bulge, \cite{1975PASJ...27..533M} disk and the Navarro-Frenk-White \citep[NFW]{1996ApJ...462..563N} DM halo.
The concentration parameter describes the characteristic inner scale radius of the DM halo, 
and evolves with time as the galaxy accretes more and more satellites. 
It can be found by combining the Table 1 and the Eqn. 5 of \citetalias{2005ApJ...635..931B}
and we provide the value for each case in Fig.~\ref{fig:bnjallhalomass}.
Also, in the figure we display the present day (redshift, $z=0$) masses of the parent galaxy as a function of halo-centric distance $r$, 
where the total combined mass due to bulge, disk and DM halo is shown with solid lines and 
the sole contribution of the DM halo is shown with the dashed-lines.
The horizontal and vertical black solid lines demarcate the virial mass $M_\text{vir}=1.4\times10^{12}\,\Msun$ 
and the virial radius $r_\text{vir} = 289\,\kpc$ respectively, which are constants for all the simulated stellar halos. 
In the figure we see that the effect of choice of concentration on the overall mass profiles of the parent galaxy is substantial at small radii, and negligible near and past the virial radius. 

It is important to note here that while the disk of the parent galaxy is assumed to have an axisymmetric potential, 
the underlying DM halo potential is spherically symmetric and this may be of some concern as the DM halos under \LCDM\ paradigm are expected to be of triaxial nature \citep[e.g.][]{2002ApJ...574..538J,2014MNRAS.439.2863V}. 
We discuss the cases with triaxial halo in Section~\ref{sec:triaxial}.

\subsection{Simulated stellar halos with artificial accretion history}
Besides the \LCDM\ halos, we also utilize additional six simulated stellar halos obtained from \citetalias{2008ApJ...689..936J}, which we refer to as artificial halos.
These halos are also simulated in the exact same way as the \LCDM\ halos except that these halos have artificial accretion history, that is they are constructed by collaging accretion events from a library of satellites involved in creating the \LCDM\ halos described previously, but chosen to acquire the following properties:
\begin{itemize} 
\renewcommand{\labelitemi}{\scriptsize$\blacksquare$} 
\item \emph{rad}: a radial halo built from events predominantly on radial orbits, i.e., ratio of the angular momenta of the orbit to a circular orbit of same energy, $\epsilon<0.2$
\item \emph{circ}: a circular halo built from events predominantly on circular orbits, i.e., ratio of the angular momenta of the orbit to a circular orbit of same energy, $\epsilon>0.7$
\item \emph{old/ancient}: a old halo built from events entirely accreted, the time since accretion, more than 11 Gyr ago,
\item \emph{young/recent}: a young halo built from events entirely accreted, the time since accretion, less than 8 Gyr ago.   
\item \emph{highl}: a high in luminosity halo built from events that were more luminous than $10^{7} L_\odot$, and 
\item \emph{lowl}: a low in luminosity halo built from events that were less luminous than $10^{7} L_\odot$.
\end{itemize}
These halos possess roughly $10^9 L_\odot$ stellar tracers.
The artificial halos are not cosmologically motivated as they are constructed 
from a collage of different satellites that have been
evolved in different host potential having different
concentration parameter $c$. 
Hence, the intrinsic mass profiles of the resultant
artificial halos cannot be properly defined. 
Therefore, we only use these halos to study the sensitivity of the properties of substructures in the halo to accretion history.

All of the above \LCDM\ as well as artificial halos are built up from the disruption of >300 satellites.
Among these satellites, only the 15 largest satellites account for $75-90\%$ of the luminosity of the halo.
As such, we are only interested in the tracer population dispersed from disrupted satellites 
that orbit and hence, trace the underlying potential of the parent galaxy.
Therefore, we exclude the particles that are still bound to the host satellites as they are not in equilibrium with the parent galaxy. 

In Fig.~\ref{fig:rvr} we show the radial velocity distributions of the stellar tracer population of the simulated stellar halos as a function of halo-centric distance $r/\kpc\in[1,100]$. 
In the bottom panel where we show the artificial halos, we see that the tracers of the \emph{young} halo are relatively less relaxed compare to 
the \emph{old}, \emph{rad} and \emph{lowl} halos, which are better phase-mixed and is as expected. 
In the top panel of the figure where we show the 11 \LCDM\ halos demonstrate diverse distribution of sub-structures, akin to the distinct accretion history of the halos.
The effective number of star particles in these simulated stellar halos ranges between $2-8 \times 10^4$, exception to this are the \emph{rad}, \emph{circ} and \emph{lowl} halos which have $10^5$, $10^5$ and $6\times10^5$ 
particles respectively.

\section{Dynamical Mass: Jeans Formalism}\label{sec:jeans}
\begin{figure*}
  \centering
  \includegraphics[width=2.1\columnwidth]{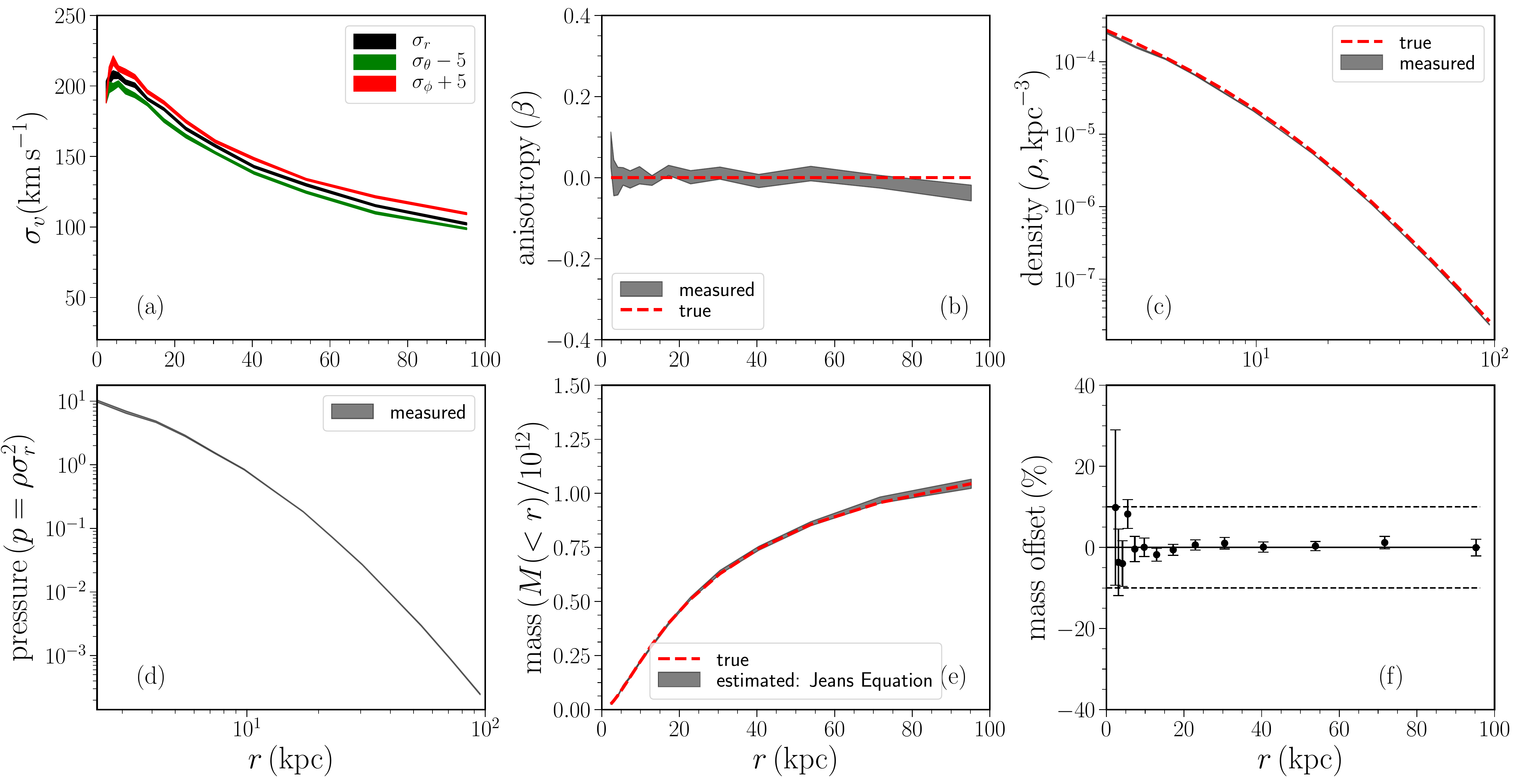}
    \caption{Jeans formalism implemented on a synthetic data sampled from the Hernquist distribution. Panels show (a) radial (red line), angular (green line) and azimuthal (black line) velocity dispersions, (b) velocity anisotropy (grey band: measured, red dashed-line: the input profile), (c) number density (grey band: measured, red dashed-line: the input profile), (d) measured radial pressure (e) cumulative mass (grey band: reconstructed from the Jeans equation, red dashed-line: the input profile), and (f) percentage error in the mass measurement (black dots) with dashed lines at $\pm10\%$ as a guide.}
   \label{fig:test_withHQ}
\end{figure*}
\subsection{Formulary}
The \cite{1915MNRAS..76...70J} equation for a pressure-supported, collision-less
and spherically symmetric dynamical system in equilibrium is customarily expressed in spherical polar coordinates as follows:
\begin{equation}\label{eqn:jeanseqn}
  M_{\textrm{Jeans}}(<r) = -\frac{r\,\rsigma^2}{G} \left( \frac{\text{d} \ln \rho}{\text{d}\ln r}  + \frac{\text{d} \ln \rsigma^2}{\text{d}\ln r} + 2\,\beta \right)
\end{equation}
\citep{2008ApJ...684.1143X,2012MNRAS.424L..44D,2012ApJ...761...98K,2013MNRAS.432.3361R,2015ApJ...813...89K}.
When number density $\rho(r)$, radial velocity dispersion $\sigma_r(r)$ and velocity anisotropy $\beta(r)$ runs of the dynamical tracer population are known, 
one can substitute them in the Jeans equation to derive the underlying mass $M_{\textrm{Jeans}}(<r)$ (or $M$ for conciseness) distribution of the system.
Here, $r$ stands for a distance from the centre of the system,
and the velocity anisotropy is expressed as
\[\beta=1 - \frac{\thsigma^2 + \phsigma^2}{2 \rsigma^2},\]
where $\thsigma$ and $\phsigma$ are angular and azimuthal velocity dispersions of the tracer population. 
The $\beta$ parameter can have values between $[-\infty, 1]$, 
where $\beta<0, \beta =0, \,\text{and}\, \beta>0$ signify tangential, isotropic and radial system.

We rearrange equation~\ref{eqn:jeanseqn} and express it in terms of the radial pressure term $(p=\rho\sigma_r^2)$ as follow:
\begin{equation}\label{eqn:p_jeanseqn}
M(<r) = -\frac{r}{G} \left[ \sigma_r^2 \left( \frac{\text{d} \ln p}{\text{d} \ln r} + 2 \right) - (\sigma_\theta^2 + \sigma_\phi^2)  \right]
\end{equation}
Compared to equation~\ref{eqn:jeanseqn}, equation~\ref{eqn:p_jeanseqn} has one less differentiation operation and also, it does not contain the $\beta$ term explicitly, 
which makes the analysis simpler.
We numerically compute the terms in the right side of the equation in concentric radial shells. 
The density $\rho_i$ of the $i_{\text{th}}$ bin between radii $r_{i}$ and $r_{i+1}$  
is calculated directly from the frequency of the corresponding bin $n_i$ using
\begin{equation}\label{eqn:density}
\rho_i = \frac{n_i}{(4\pi/3) (r_{i+1}^3 - r_{i}^3) \sum_i n_i}.
\end{equation}
Here, we have rescaled the density by the total number of tracers ($\sum_i n_i$)
and this is to facilitate comparison amongst the simulated halos with varying number of tracers.
The scaling has no effect in the final mass measurement as the Jeans equation 
only demands for the logarithmic slope of the density distribution and not its normalisation.
In our case, where star particles are essentially the accreted DM tracer particles with individual luminosity weight ($w_i$),
we replace the frequency $n_i$ with the effective sample size = $(\sum_i w_i)^2/\sum_i w_i^2$.
Similarly, in this case we calculate the weighted velocity dispersions using  
\begin{equation}\label{eqn:veldisp}
\sigma_w = \sqrt{\frac{\sum_i w_i v_i^2}{\sum_i w_i} - \left( \frac{\sum_i w_i v_i}{\sum_i w_i} \right)^2},
\end{equation} 
where, $v_i$ represents $v_r/v_\theta/v_\phi$, spherical polar component of the velocity vector of the $i_{\text{th}}$ star particle.

Irrespective of the chosen form of the Jeans equation, there are some sources of noise, which are inevitable as they akin to physical processes. 
For example, the number of dynamical tracers drop sharply at large radii resulting large scatter in $\rho$ and $p$ estimates.  
Additionally, dispersed (not yet fully phase-mixed) substructures are ubiquitous feature of the \LCDM\ halos that can be present at different radius.
In such radial shells velocity dispersions and density measurements are likely to be biased and dominated by the substructures.
Consequently, this can bias the slope of the radial pressure ($\text{d} \ln p/\text{d} \ln r$) as well.
Also, in the very inner $r\lesssim1$ kpc region of the simulated stellar halo there is usually a sharp rise and then fall in the velocity dispersion profiles.
The sudden change in the kinematics for a small change in radius makes the pressure-slope ($\text{d}\ln p/\text{d}\ln r$) poorly measured. 
Moreover, in reality this region is completely dominated by the bulge and the disc, where it is not relevant to consider finding the potential from the halo tracers.
For these reasons we restrict our study to $1<r/\text{kpc}<100$.

Finally, we estimate the scatter around our all measurements using the bootstrap scheme, which is implemented by constructing 100 sub-samples (with replacement) of the dataset.
We consider the mean and standard deviation of the bootstrapped measurements at each bin as our best measurement and associated uncertainties of the relevant physical quantities respectively.
Moreover, to quantify the bias in the mass measurement we define
\begin{equation}\label{eqn:massoffset}
{\rm mass\ offset (in\ per\ centage)} = 100 \times \frac{M - M_\text{True}}{M_\text{True},}
\end{equation}
where $M_\text{True}$ denotes the intrinsic mass profile of the parent galaxy (shown in Fig.~\ref{fig:bnjallhalomass})
whereas $M$ represents the galaxy mass estimated from the Jeans formalism.

\subsection{Tests with a featureless simple model}\label{sec:hqtest}
To investigate inherent biases in our Jeans formalism, first we test the scheme on toy data sampled from an ergodic \cite{1990ApJ...356..359H} model
with scale-length $a=15$~kpc, total mass = $1.4 \times 10^{12}~\Msun$, and isotropic velocity distribution ($\beta=0$).
By nature the data is smooth, i.e., devoid of any sub-structures or tidal features, and therefore allows us to understand, if any, intrinsic bias in our scheme.
To facilitate comparison to our main results with simulated stellar halos, we generate a sample roughly of the same order of magnitude, i.e., $10^5$.

In our context where we are probing an order of magnitude range in distances and also, where sample size sharply declines at large distance 
the logarithmic binning in radius is more appropriate compare to a linear equal width or equal number binning schemes.
Therefore, we divide our sample into 15 logarithmically spaced radial shells, and present the key spatio-kinematics properties of the tracers in Fig.~\ref{fig:test_withHQ}.
Intrinsically all three velocity dispersion profiles ($\rsigma$, $\thsigma$ and $\phsigma$) 
of the tracer population are identical, therefore to make them visually distinct we systematically shift the angular and azimuthal dispersions by $\pm 5 \kms$ (Fig.~\ref{fig:test_withHQ} a).
In the figure thickness of all, but except red-dashed, lines presented in panels (a)-(e) show the $1\sigma$ uncertainty levels. 
Panel (e) compares the underlying mass distribution using the Jeans formalism with the intrinsic values of the system, and finally, panel (f) shows the residual in the mass measurement.
In summary, the exercise demonstrates that the spatio-kinematic runs of the test data and the underlying mass distributions of the system at $r>5~\kpc$ can be recovered well enough with negligible mass offset of $0.14^{+1.3}_{-1.9}\%$, where average dispersion is consistent with the median random uncertainties of $1.6\%$.
However, due to abrupt change in the slopes of the velocity dispersions the mass profiles 
in the inner region deteriorates.
Additionally, we also achieve similar level of accuracy for alternative cases such as when the velocity distributions of the data are assumed to be radial ($\beta=0.5$ and $0.9$) or tangential ($\beta=-0.5$ and $-1.0$).

\section{Results}\label{sec:results}
\subsection{Spatio-kinematic profiles of the simulated stellar halos}
\begin{figure*}
  \centering
  \includegraphics[width=0.925\columnwidth]{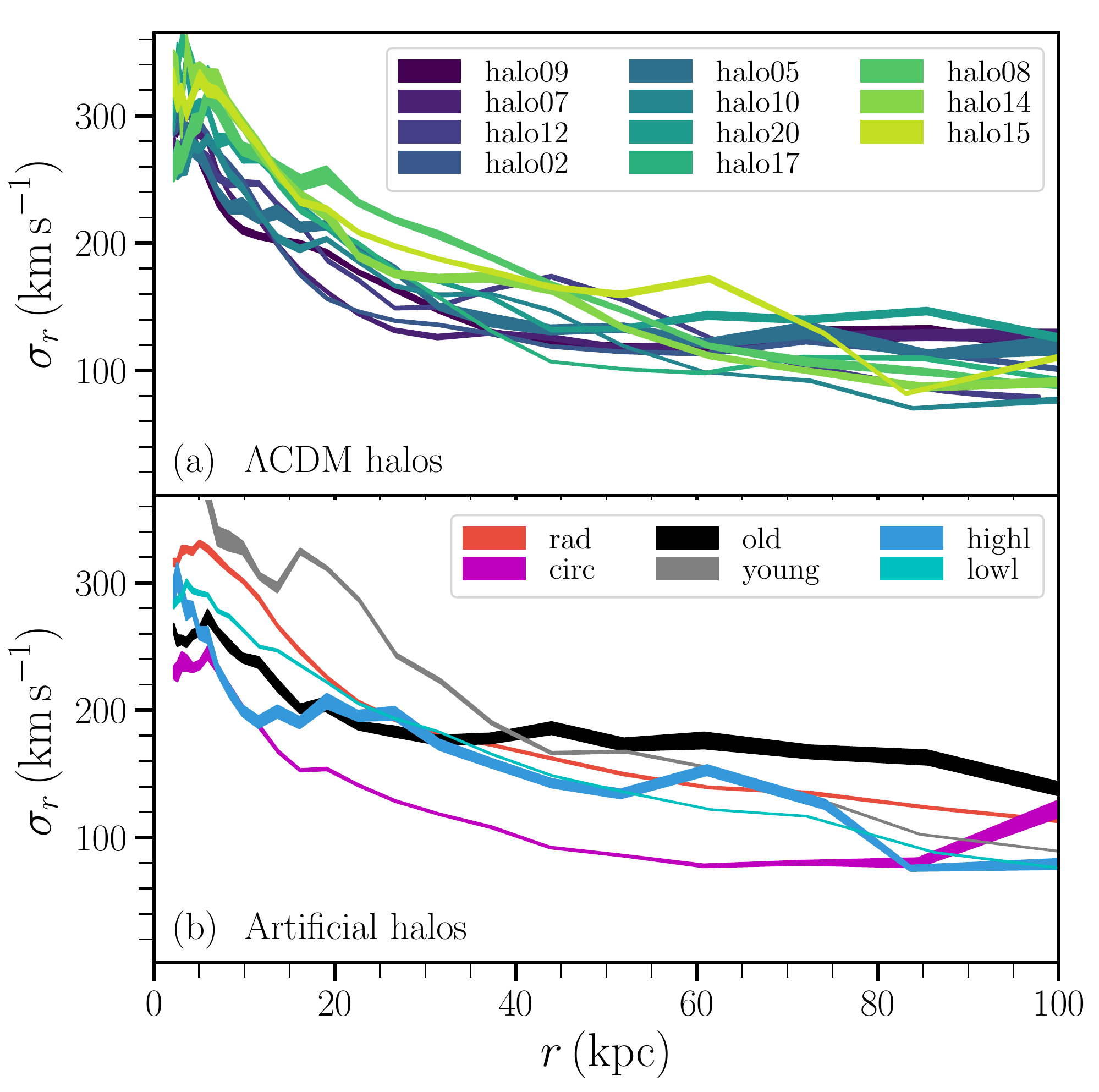}
  \includegraphics[width=0.925\columnwidth]{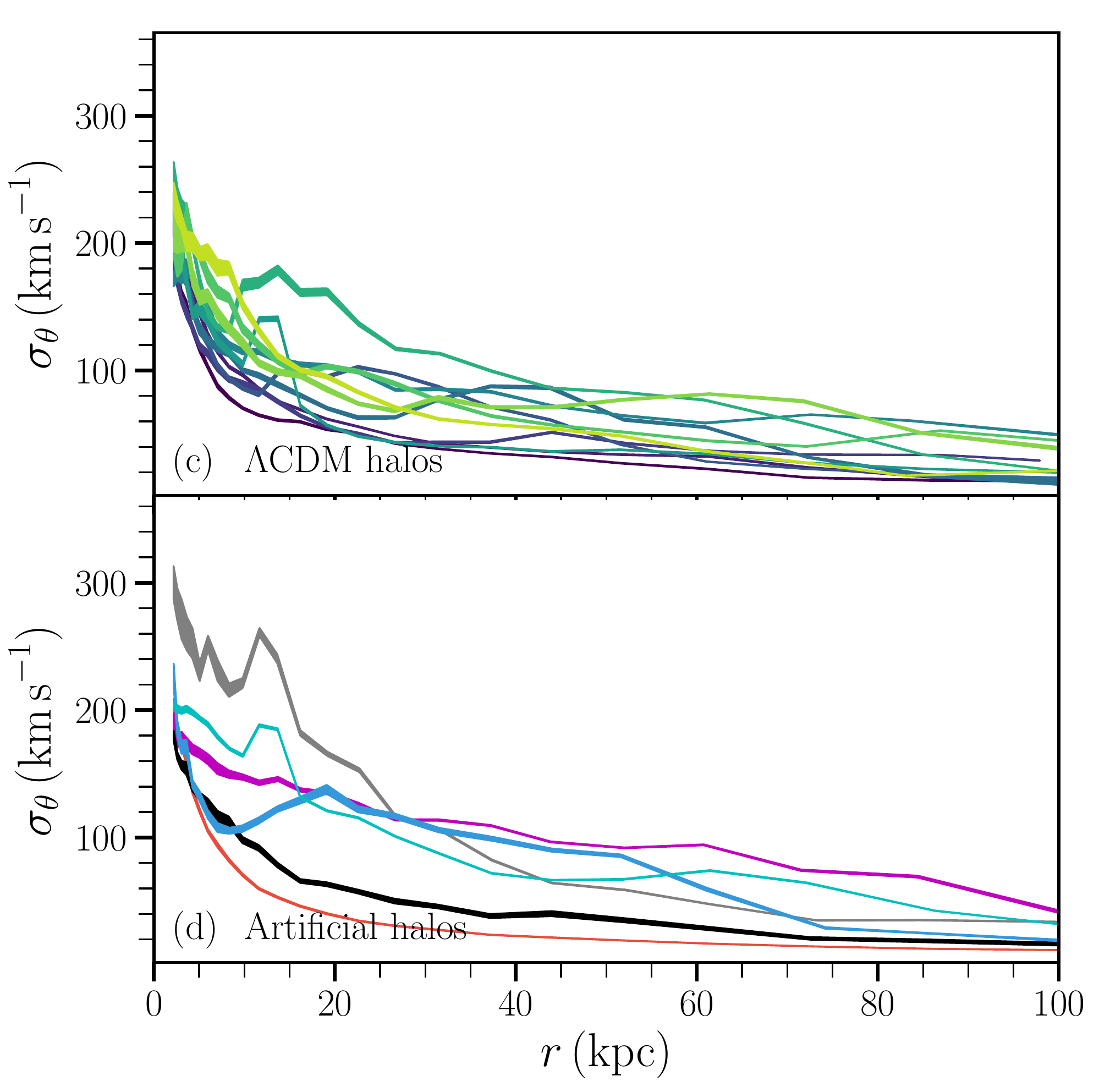}
  \includegraphics[width=0.925\columnwidth]{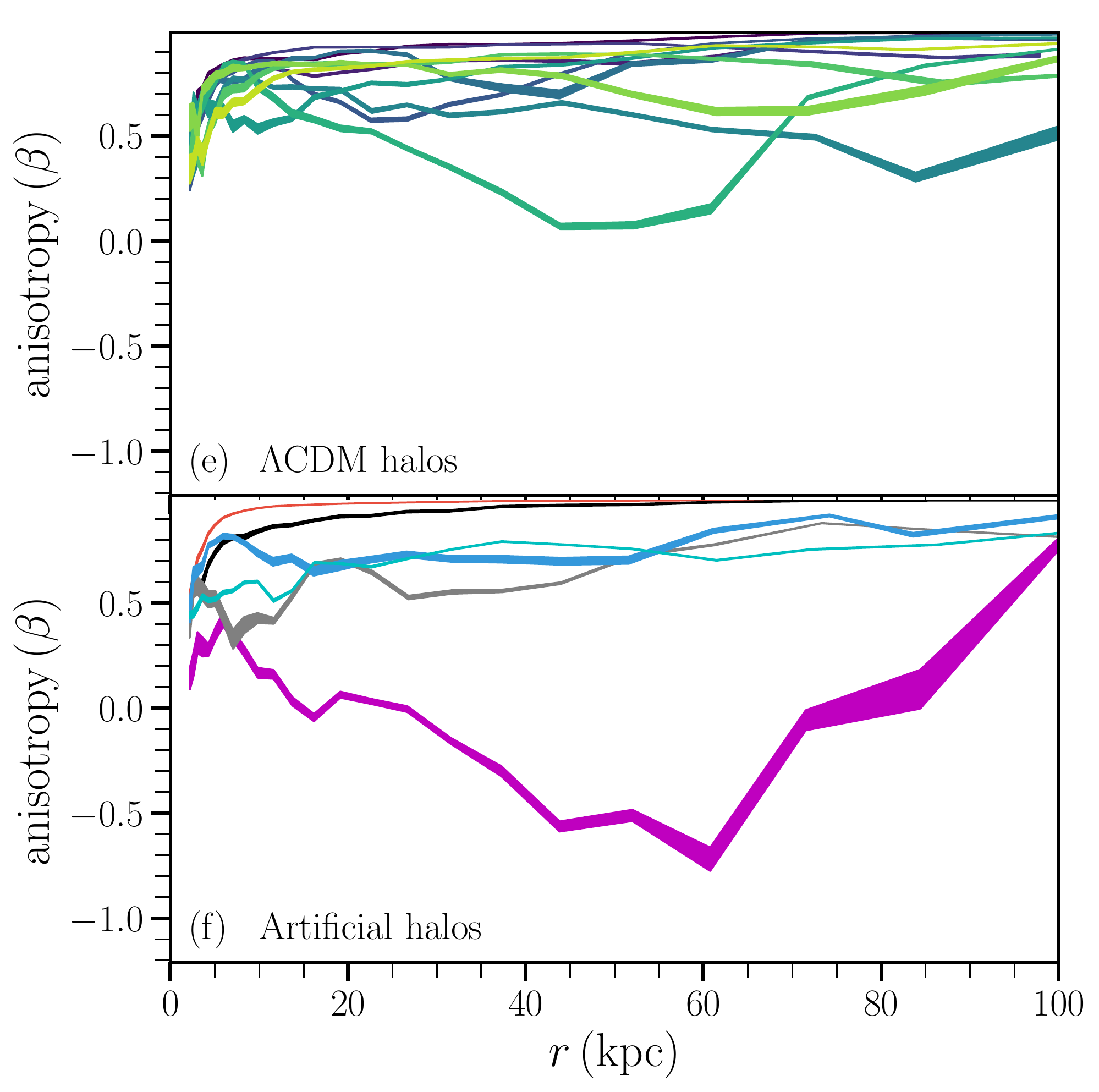}  
  \includegraphics[width=0.925\columnwidth]{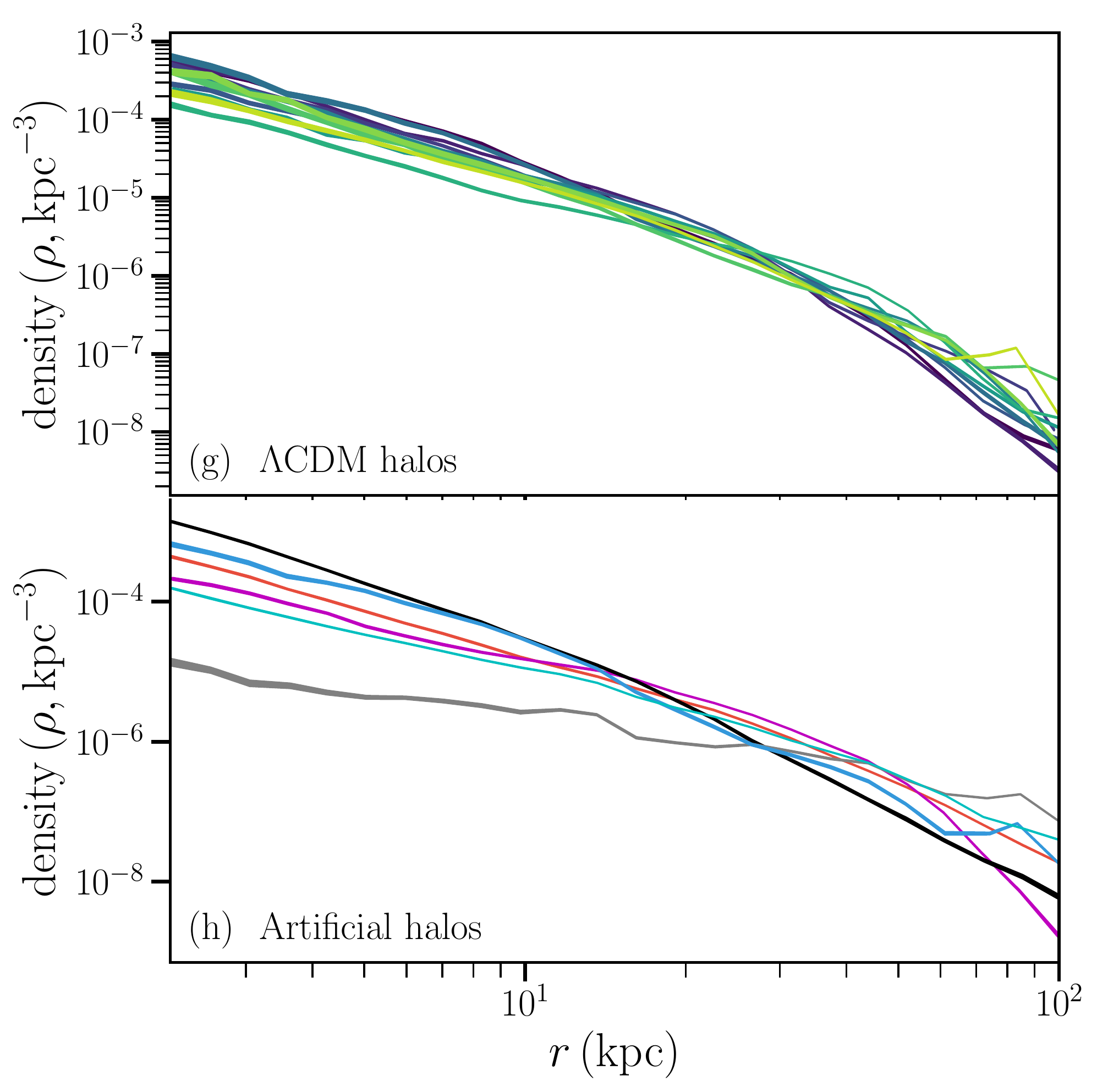}  
  \includegraphics[width=0.925\columnwidth]{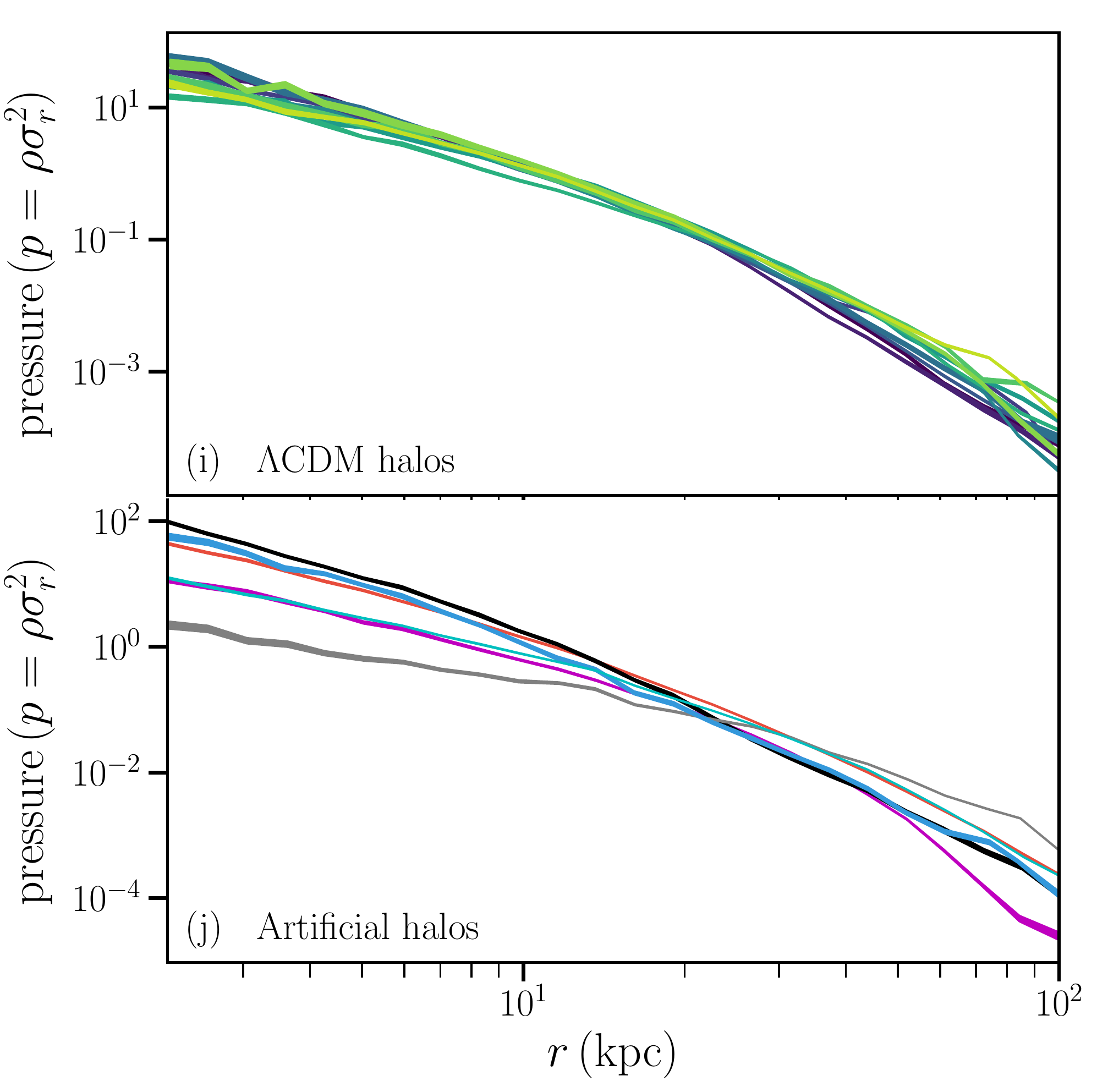}
  \includegraphics[width=0.925\columnwidth]{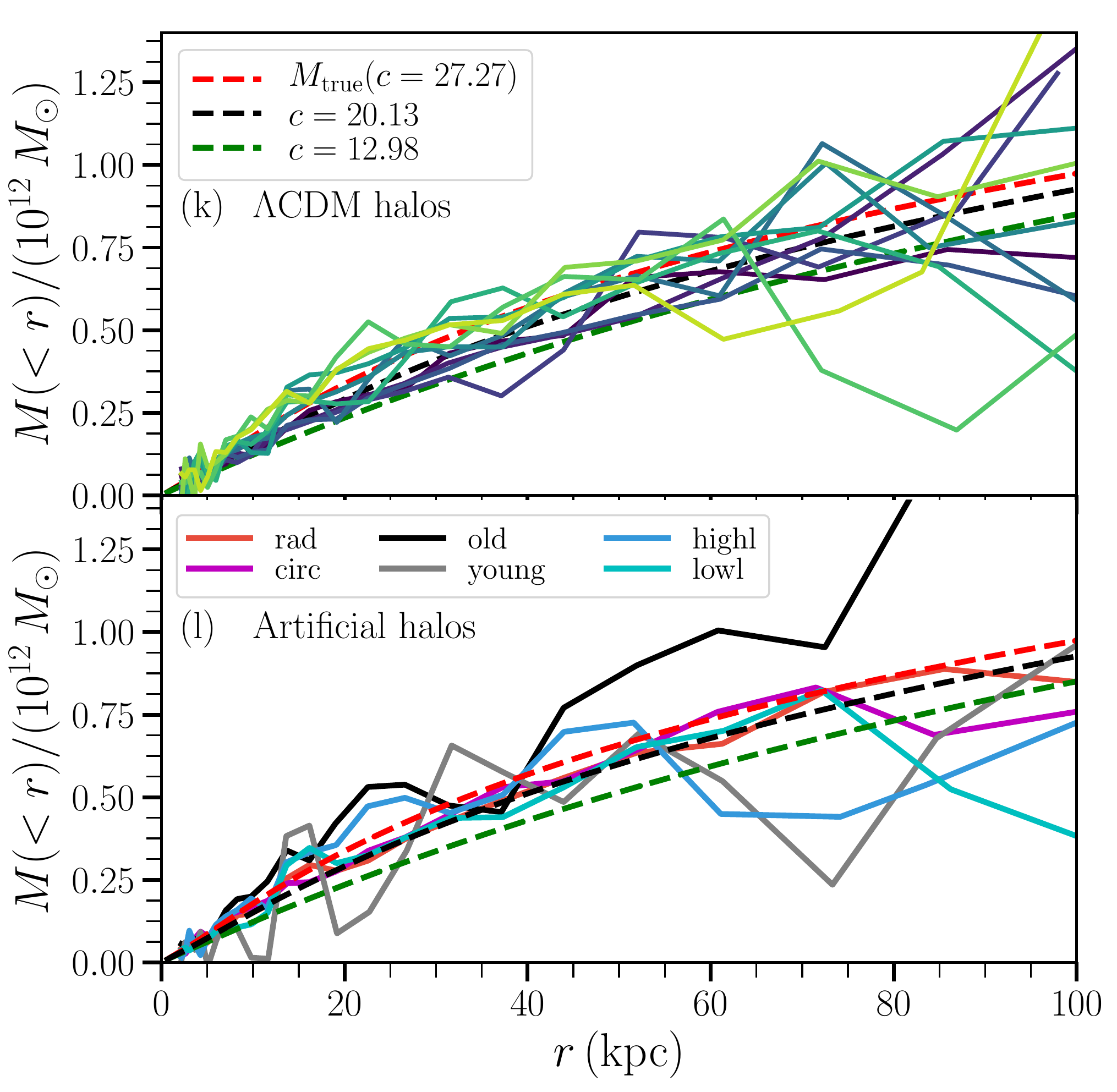}
  \caption{Key spatio-kinematic properties of the tracer population of the simulated stellar halos and inferred mass distribution of the parent galaxy taken from \protect\citetalias{2005ApJ...635..931B} and \protect\citetalias{2008ApJ...689..936J} simulations.}
  \label{fig:profiles_wt}
\end{figure*}

In Fig.~\ref{fig:profiles_wt} we demonstrate the key spatial and kinematic properties of the tracer populations of the 11 \LCDM\ (top panels) and 6 artificial (bottom panels) halos binned respectively in 25 concentric radial shells.
As labelled in the figure, each solid line of different colour represents different halo. 
Panels (a) and (b) show the measured radial velocity dispersion $\rsigma(r)$, panels (c) and (d) show the measured angular\footnote{to save the space we do not show the azimuthal velocity dispersion $\phsigma$, which generally has runs identical to $\thsigma$.} velocity dispersion $\thsigma(r)$, and panels (e) and (f) show the corresponding velocity anisotropy $\beta(r)$ profiles of the tracer populations.
Similarly, panels (g and h) and (i and j) show the measured number density and pressure distributions of the tracer population.
Finally, substituting the above measured tracer properties into the Jeans formalism (equation~\ref{eqn:p_jeanseqn}) we estimate the mass profiles of the parent galaxy, 
which are shown with solid lines in panels (k) and (l).
For an easy comparison, in panels (k) and (l) we also over-plot the inherent mass profiles of the parent galaxy assuming maximum (red dashed-line), minimum (green dashed-line) and average (black dashed-line) values of concentration adopted from Fig.~\ref{fig:bnjallhalomass}.
The width of the lines in panels (a-j) depict the scatter around the relation obtained from bootstrapping whereas to avoid further cluttering we only show mean relations in panels (k) and (l).
There are a few trends we note in Fig.~\ref{fig:profiles_wt}. 
For example, both the $\rsigma$ and $\thsigma$ (also $\phsigma$, not shown in the figure) attain highest value for small $r$ and vice-versa, which turn-over at $r \simeq 5$ kpc.
Also, all of the halos except the {\it circ} have predominantly radial orbits i.e. $\beta>0$ (see panels e and f).
Furthermore, it can be visually attested that the logarithmic density distributions $\rho(r)$ of the tracer particles seem to follow a double power-law with a break at $r \sim20$ kpc. 

\subsection{Mass reconstruction from tracer population}
\begin{figure*}
  \centering
  \includegraphics[width=2.1\columnwidth]{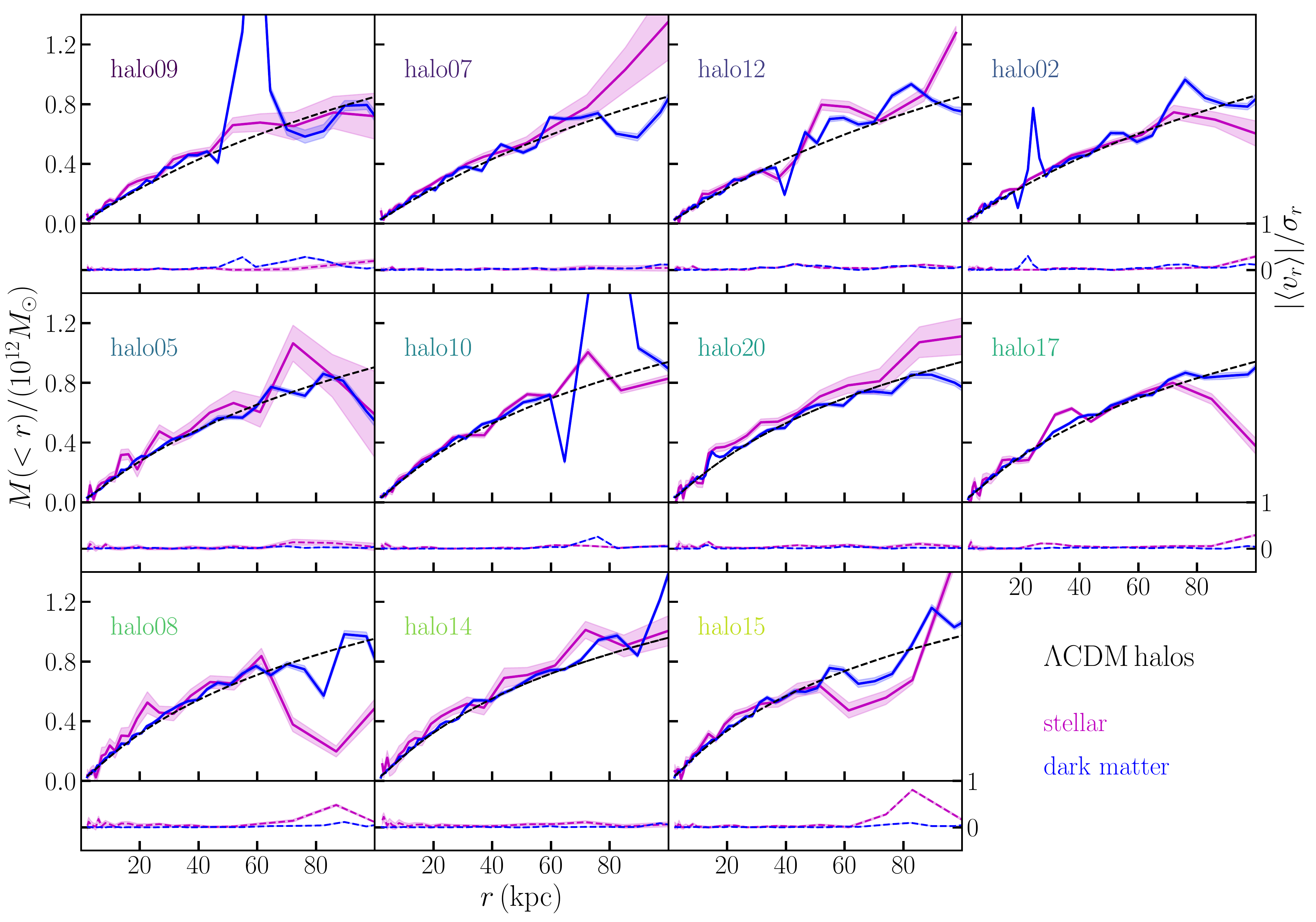}
  \includegraphics[width=2.1\columnwidth]{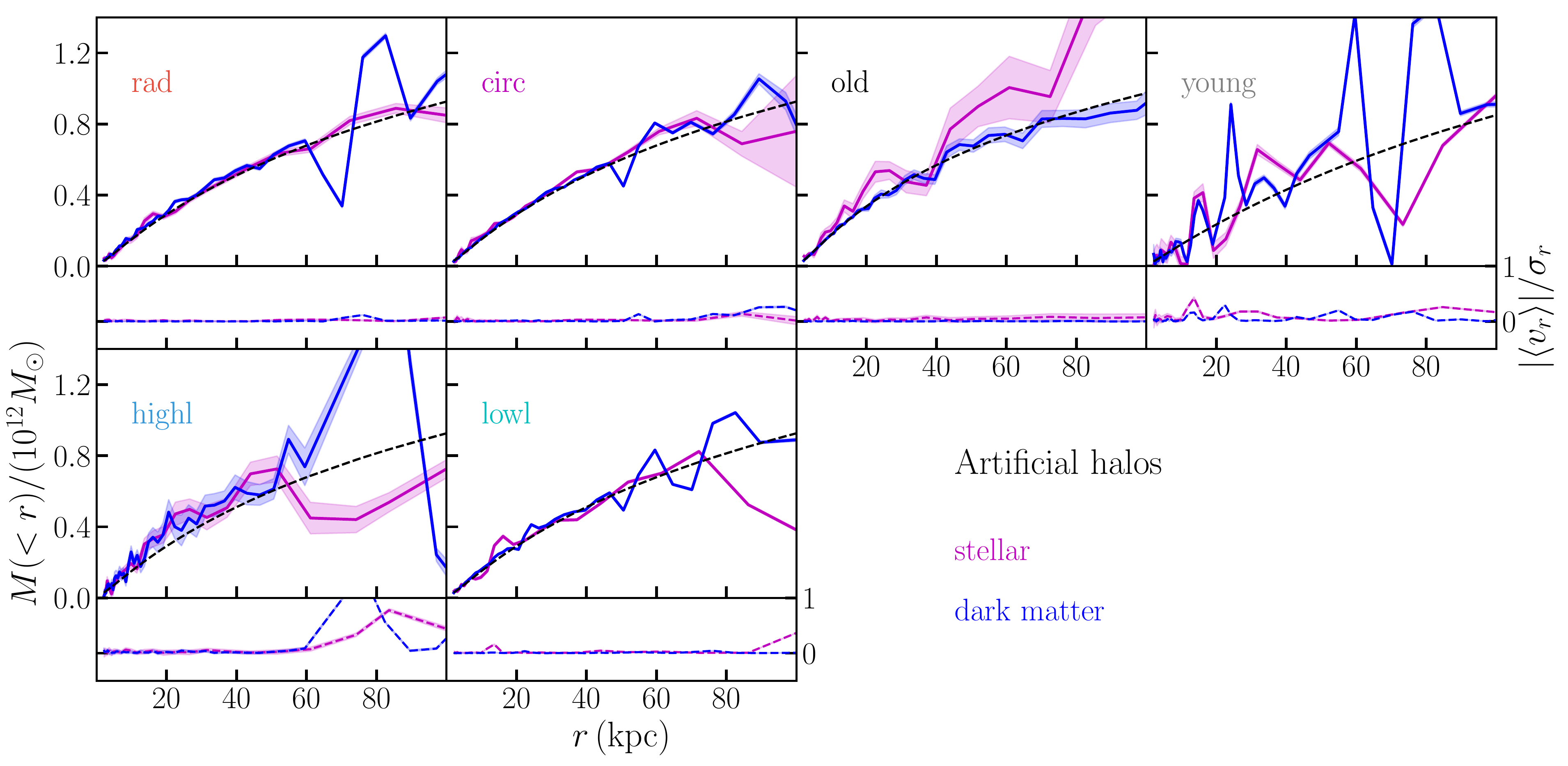}
  \caption{Mass profiles of the parent galaxy obtained from the Jeans formalism using the spatio-kinematic profiles of the tracer populations. Black dashed lines show the intrinsic mass profile of the parent galaxy whereas magenta and blue solid solid lines are estimated masses of the galaxy when stellar and accreted DM particles are used as dynamical tracers respectively. The magenta and blue dashed lines shown in small panels at ordinates $\simeq0$ are respective measurements of $|\langle v_r \rangle|/\sigma_r$. The bands of corresponding colours around the blue and magenta lines show the associated uncertainties obtained from the bootstrapping.}
\label{fig:massprofile}
\end{figure*}
\begin{figure*}
  \centering 
  \includegraphics[width=1\columnwidth]{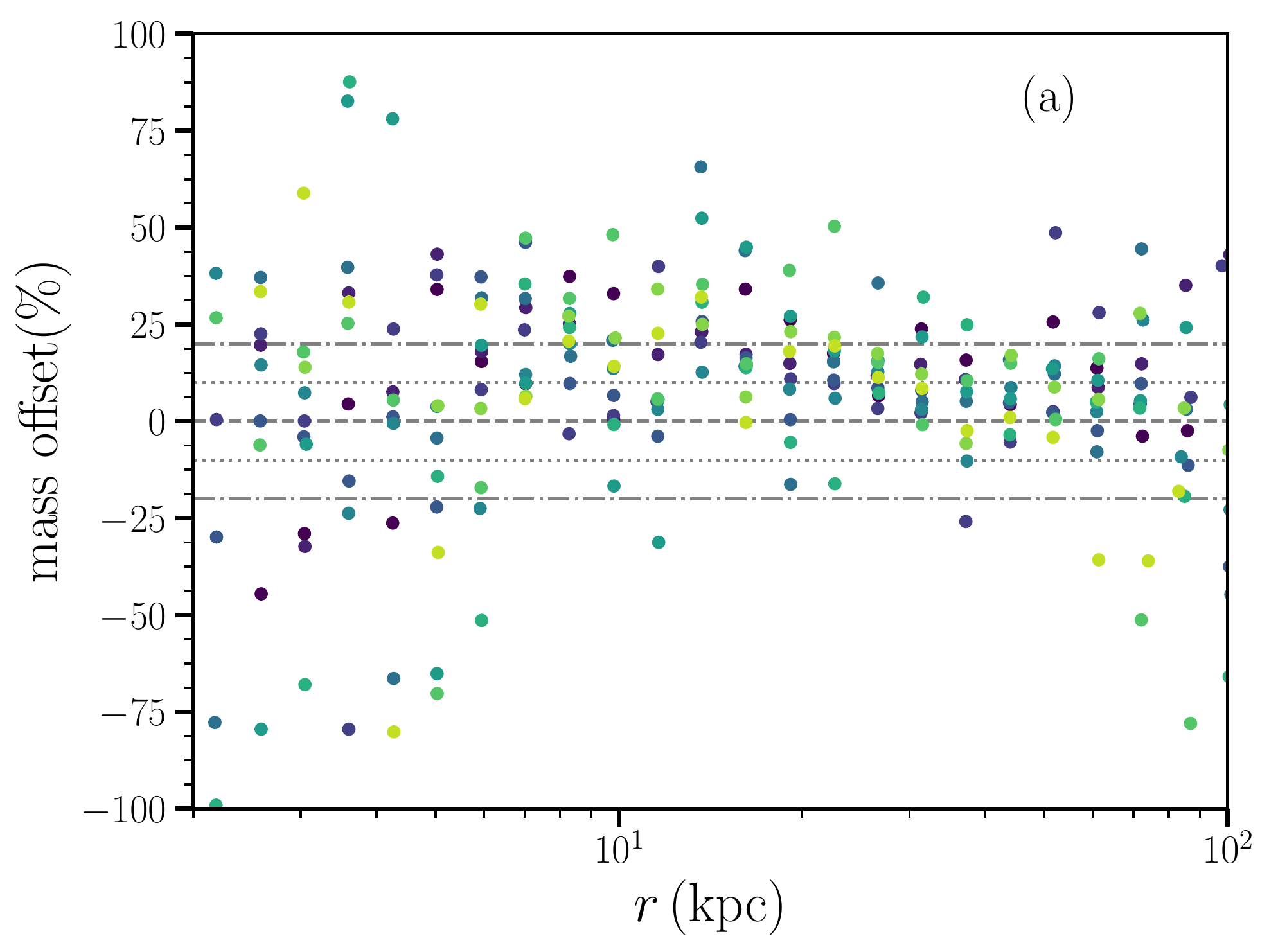}
  \includegraphics[width=1\columnwidth]{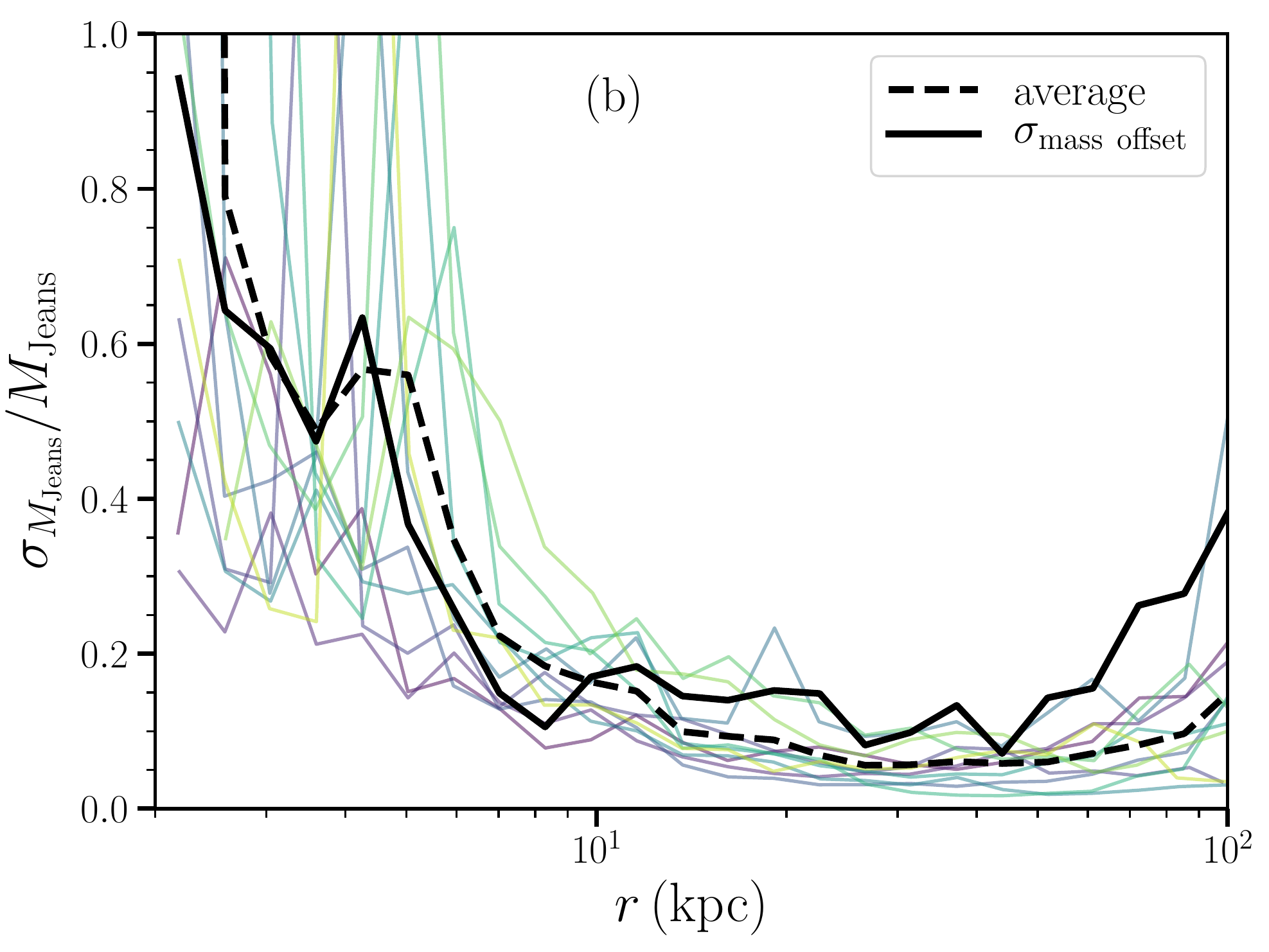}
  \includegraphics[width=1\columnwidth]{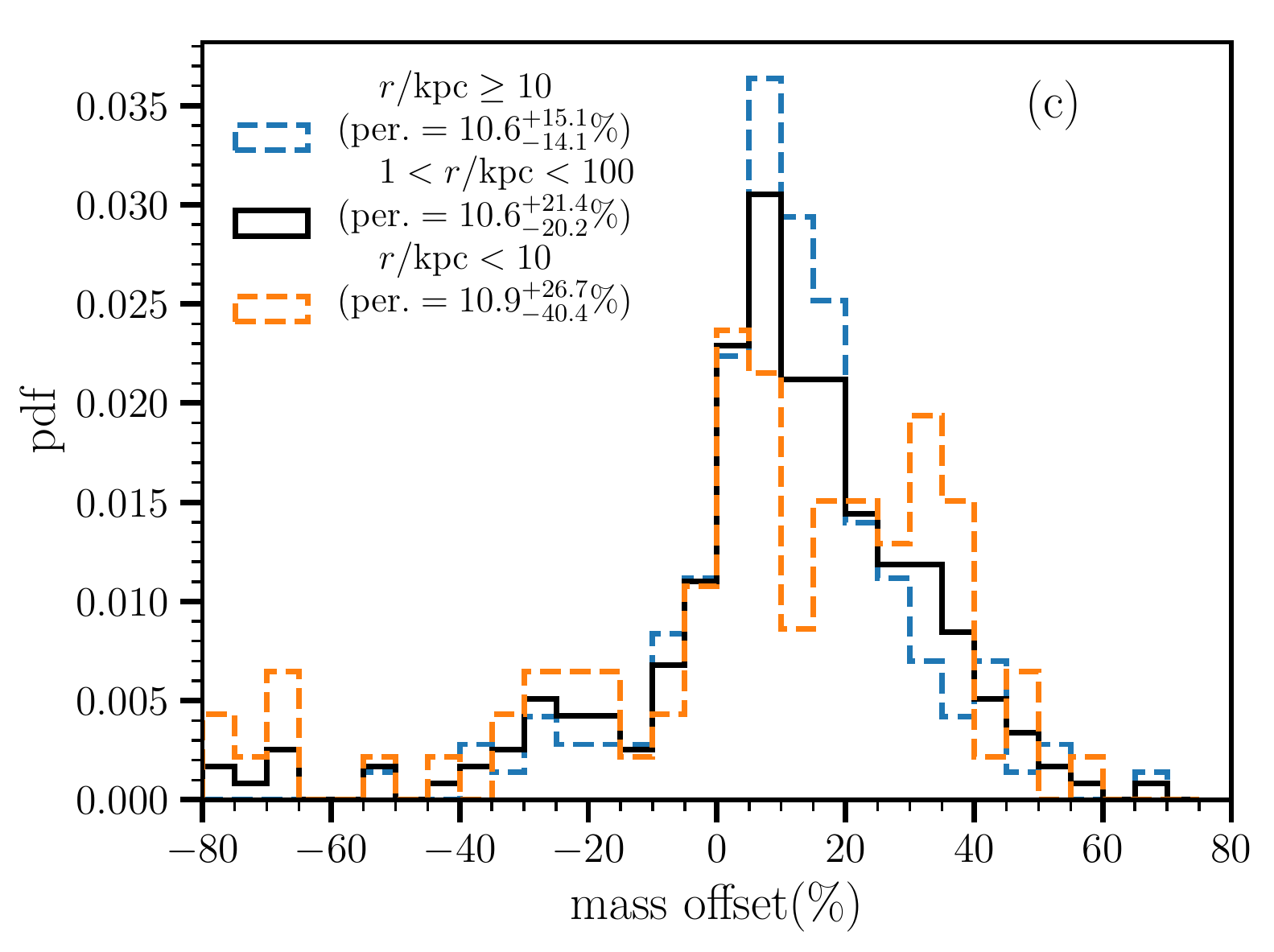}
  \includegraphics[width=1\columnwidth]{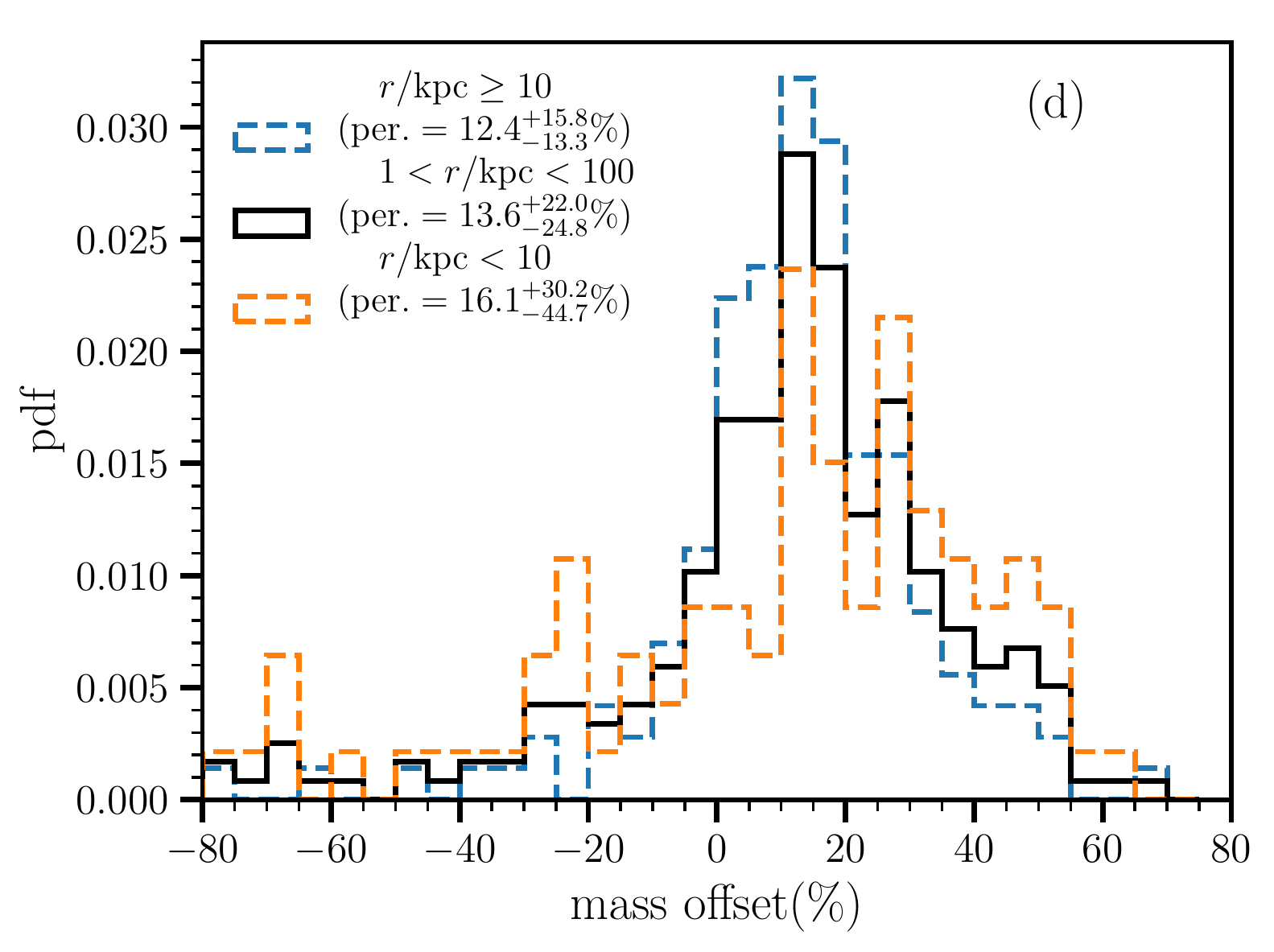}  
  \caption{Error analysis of the estimated masses of the parent galaxy using the stellar tracer population of the 11 \LCDM\ halos. Panel (a) shows the percentage mass offset and panel (b) 
  shows the random uncertainties in mass measurements, both as a function of distance, 
  where different colours represent different halos as labelled in Fig.\ref{fig:bnjallhalomass} and for the aesthetic reason we have not shown the uncertainty bands around these trends. 
  Panels (c) and (d) show distributions of the mass offset for cases where we respectively use standard deviation and root-mean square velocity as a measure for the velocity dispersion.}
  \label{fig:lcdm_mpe}
\end{figure*}
For clarity, in Fig.~\ref{fig:massprofile} we recast the mass profiles of the parent galaxy of 11 \LCDM\ (top panels) and 
the 6 artificial (bottom panels) halos, taken originally from panels (k) and (l) of Fig.~\ref{fig:profiles_wt}.
The magenta bands with solid lines show the mass profiles of the parent galaxy 
reconstructed from the Jeans equation using the spatio-kinematic profiles of the stellar tracer populations.
The bands around the lines show the bootstrapped uncertainties around the mean measurements.
Furthermore, the intrinsic mass distributions of the parent galaxy is shown with the black dashed-line. 
As mentioned earlier in Section ~\ref{sec:data}, the intrinsic mass profiles of the artificial halos 
are not known as they are constructed from a mixture of satellites that have been evolved in different host potential.
Therefore, in the case of artificial halos we only use the black dashed-lines as a rough guide for the purpose of a qualitative assessment, 
which are derived assuming the concentration of the artificial halos to be an average concentration of the satellites that make up these halos.
The spikes in the mass profiles are far more prominent than the uncertainties in the measured values, therefore
statistical noise as a potential cause for the spikes can be ruled out. We now investigate the cause of these spikes.

We expect that the mass distributions of the parent galaxy monotonically increase as a function of radius.
However, this is not strictly the case for our estimated mass profiles, again magenta lines shown in Fig.~\ref{fig:massprofile}.
At many places the mass profiles are bumpy (e.g. {\it halo 09} at $r\sim60~\kpc$, {\it halo 10} at $r\sim70~\kpc$) 
and in some cases the cumulative mass dips (e.g. {\it halo 08} at $r>60~\kpc$, {\it highl} halo at $r>50~\kpc$), which is unphysical.
Similarly, the mass profiles of the {\it highl} halo at $r\gtrsim60~\kpc$ and {\it young} halo at all radius are also bumpy. 
This discrepant output of our scheme can be understood in conjunction with the radial phase-space diagram shown in Fig.~\ref{fig:rvr}.
It is clear from this figure that at some intermittent radius satellites are just dispersing (say, {\it halo 15} at $r\sim80~ \kpc$, {\it halo 09} at $r\sim60~ \kpc$ etc).
In such radial shells we will measure biased $\sigma_r$ and $\rho(r)$ and hence, 
the Jeans equation locally fails here as it demands velocity dispersions of stars evenly populating the velocity range not one biased by clumps.
The {\it young} halo (built from recent <8 Gyr ago events), which have not had enough time to fully relax suffers the worst from this scenario. 
Also, {\it highl} halo (dominated by the massive accretion events) shows significant undulations in mass profiles at large radius. 
On the contrary, halos such as {\it halo14} does not contain any dominant sub-structures and hence, result a well-behaving mass profile. 
Similarly, the halos such as {\it old} and {\it lowl}, which are expected to be better phase-mixed halos by construction, have comparatively smoother mass profiles. 

In the end, we like to validate if the erratic mass measurements at intermittent radii are due to the presence of sub-structures that are out of equilibrium. 
The mean radial velocity $v_r=0$ is a necessary condition for a system in equilibrium (although not a sufficient condition). So, $|\langle v_r \rangle|/\sigma_r$  can be used to gauge the departure from equilibrium. Note, for N stars in a bin $|\langle v_r \rangle|/(\sigma_r/\sqrt{N})$ is a better measure of the statistically significance of departure of $v_r$ from zero, as it takes the effect of Poisson noise into account. However, we choose to study $|\langle v_r \rangle|/\sigma_r$ as it is physically more meaningful and significant.  
From equation~\ref{eqn:jeanseqn} we get, 
\begin{equation}
\delta M/M \simeq 2 \delta \sigma_r/\sigma_r \simeq 2 v_r/\sigma_r, 
\end{equation}
assuming that the change in dispersion  $ \delta \sigma_r$  is of the order of $v_r$.  
Hence, $v_r/\sigma_r$  is useful to gauge if a change in $|\langle v_r \rangle|/\sigma_r$ is enough to explain a corresponding change in mass profile. 
The magenta dashed lines in all the tiny panels of Fig.~\ref{fig:massprofile} we show $|\langle v_r \rangle |/\sigma_r$ the profile.
Additionally, to check that a spike in $|\langle v_r \rangle|/\sigma_r$ is statistically significant and not caused by Poisson noise,  we show the dispersion with band around $|\langle v_r \rangle|/\sigma_r$ profile using bootstrapping.
We find that the erratic spikes seen in the mass profiles at many places, e.g.,
halo09 at $\sim55\,\kpc$, halo02 at $\sim20\,\kpc$, coincide with spikes in $|\langle v_r \rangle|/\sigma_r$ runs.
Moreover, we see that the mass reconstructions outside $r>60\,\kpc$ generally deteriorate for all the halos, which is mainly because of the paucity of mass tracers and dominance of unrelaxed sub-structures in the outskirts.

In the figure the blue solid and dashed lines are the corresponding reconstructed mass profiles of the parent galaxy when the accreted DM particles are used as a dynamical tracer. The dark matter particles are not directly observable and the discussion of this case is not useful for pragmatic reasons. However, for the completeness reason we briefly present and discuss the spatio-kinematic profiles and error analysis of this case in the Appendix~\ref{sec:dmcase}. For almost all the halos, we observe that the stellar tracers show comparatively less ridges in the reconstructed mass profiles compare to the cases when DM tracers are used. This is because the DM tracer population have large number of prominent substructures with zero luminosity, and hence zero stellar mass. 

The intermittent noise in the inferred mass distributions can be reduced by fitting a smooth parametric models to the velocity dispersions and pressure/density runs that enter the Jeans analysis.
We restrain from doing so mainly because a smooth model can not capture the impact of substructures in the mass profiles. Also, it leads to a natural question of what are the good models for the dispersion profiles, number density etc. Therefore, we decided to use the binned data directly. 

A takeaway point from the Fig.~\ref{fig:massprofile} is that in overall the shape of the mass profiles of the parent galaxy
of the simulated stellar halos (except for the flagged halos such as {\it highl}, {\it young}),
although bumpier at places where sub-structures locally dominate, can be recovered well using the Jeans analysis.
Below we provide a more detail account of the biases in our overall mass reconstruction.

In Fig.~\ref{fig:lcdm_mpe} we present the error analysis of the mass measurements of the parent galaxy using the stellar tracer populations for all 11 \LCDM\ halos.
Panel (a) shows the mass offsets (defined in equation~\ref{eqn:massoffset})  in the estimated mass $M_{\rm Jeans}$ compared to the intrinsic masses as a function of $r$.
Similarly, in panel (b) we show the fractional uncertainty ($\sigma_{M}/M$) on the estimated mass measured using bootstrapping as a function of $r$.
Different colours of the circles or lines in panels (a) and (b) represents different halos, the labellings consistent with Fig.~\ref{fig:bnjallhalomass}.
The average of the coloured lines in panel (b) is shown with the black dashed-line whereas the black solid line is the dispersion profile of the mass offsets obtained from panel (a).

The mass offset is due to two sources, a) random uncertainty in the estimator, which is mainly due to Poisson noise and b) the tracers not being in dynamical equilibrium with the potential in which they are orbiting. 
We can estimate the random uncertainty by bootstrapping and this is given by $\sigma_{M}$. 
We label the dispersion due to non-equilibrium effects by $\sigma_{\rm noneq}$. 
The dispersion in mass offset can then be written as 
\begin{equation}\label{eqn:sig_massoffset}
\left(\frac{\sigma_{\rm mass\ offset}}{M}\right)^2= \left(\frac{\sigma_{M}}{M}\right)^2+ \left(\frac{\sigma_{\rm noneq}}{M}\right)^2
\end{equation}
If we consider the standard error of mean around the average $\sigma_{\rm mass\ offset}/M$ relation in Fig.~\ref{fig:lcdm_mpe} (b), the solid black line coincides within the $2\sigma$ confidence interval of the mean relation in the inner $r<10~\kpc$. The two relations tracing each other in this regime means that the dispersion in mass offset is dominated by random uncertainty due to Poisson noise
and can be reduced by increasing the sample size of the tracers. Moreover, the large uncertainty in the inner $r<10~\kpc$ is of the least concern to us as we are mainly interested in measuring mass for $r\gtrsim10~\kpc$. However, for $r\gtrsim10~\kpc$, the dispersion in mass offset is consistently larger than the random uncertainty and the difference keeps on increasing with increase of $r$. This means that dispersion is dominated by non-equilibrium effects instead of random uncertainty and this dispersion cannot be reduced by increasing the sample size or the precision of observables.

Finally, to quantify the bias and dispersion in the mass measurement we present the distributions of the mass offsets in panel (c), where blue, orange and black histograms show the cases with $r\geqslant10~\kpc$, $r<10~\kpc$ and $r/\kpc \in [1,100]$ respectively. 
The median and the $16^{\rm th}$ and $84^{\rm th}$ percentile ranges are shown in the panels (denoted by per.).  
In essence, from panel (c) we conclude that in reconstructing the mass profile of the parent galaxy using the stellar halo tracer populations within $r/\kpc\in[10,100]$ the bias is $\sim10\%$ and the dispersion is $\sim14\%$.
When we include the results from the inner $r<10~\kpc$ as well, the bias remains the same whereas dispersion marginally increases to $\sim20\%$. 
Additionally, we also confirm that the bias in the mass measurements are not introduced due to the consideration of mean motion while calculating the velocity dispersion in Equation~\ref{eqn:veldisp}. 
Importantly from Fig.~\ref{fig:lcdm_mpe}(b) we measure that the random error in mass measurement in the case of $r\gtrsim10~\kpc$ is $\sim7\%$ that is approximately half of the dispersion ($\sim14\%$). 
Substituting, this value in equation~\ref{eqn:sig_massoffset} we measure the contribution of non-equilibrium effect to be $12\%$.
To investigate this we repeat our analysis with root-mean square velocity as a measure for the velocity dispersion and we observe effectively similar bias in mass measurements as 
demonstrated in panel (d) of the figure.

\subsection{Deviation from spherical symmetry}\label{sec:triaxial}
\begin{figure}
    \centering 
       \includegraphics[width=1\columnwidth]{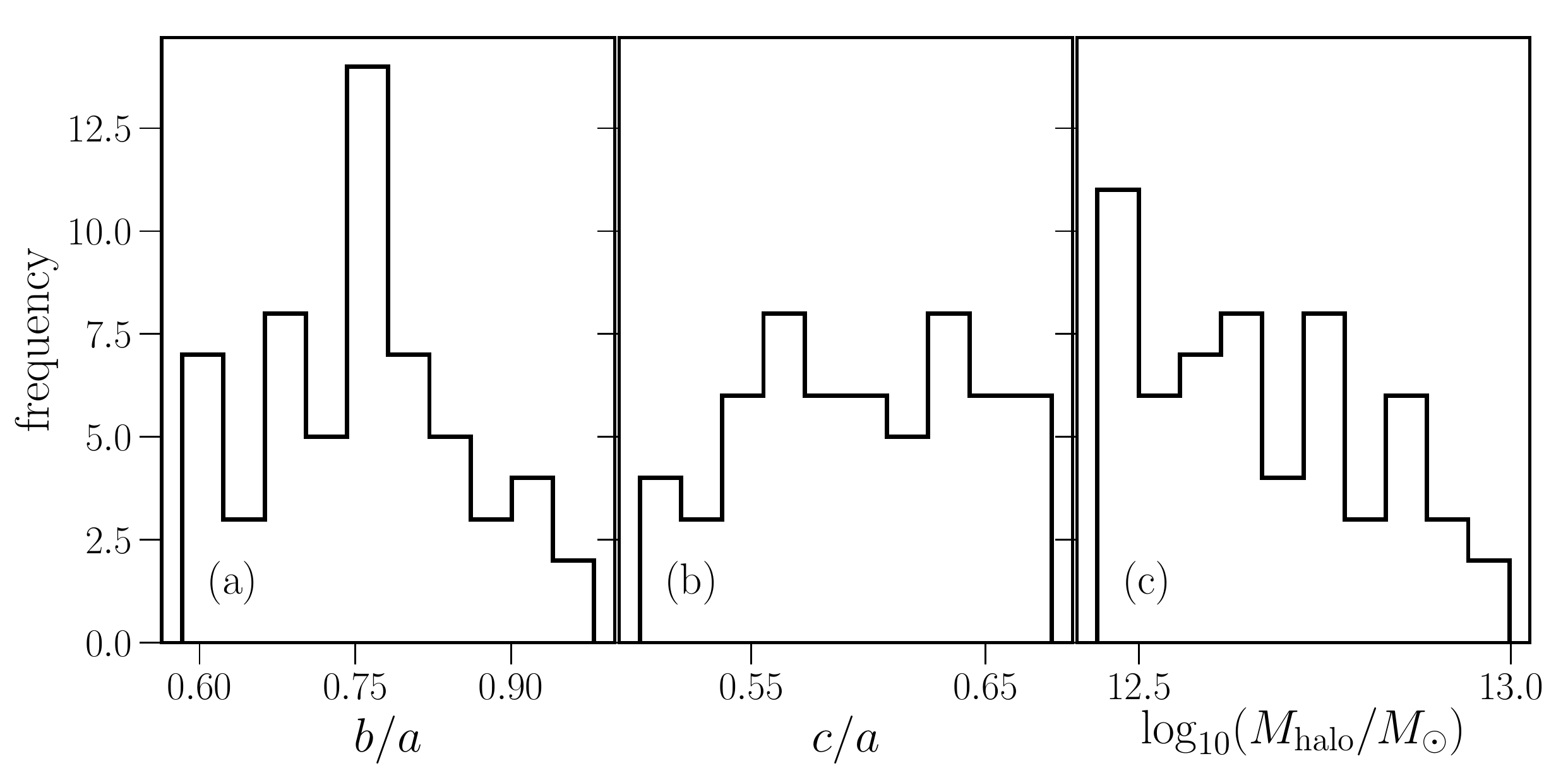}
    \caption{Distributions of (a) intermediate/major axis ratio (b) minor/major axis ratio,
             and (c) halo mass in logarithmic scale for triaxial DM halos taken from the \surfs\ simulation.}
    \label{fig:triaxialprops}
 \end{figure}
\begin{figure*}
    \centering 
        \includegraphics[width=2.1\columnwidth]{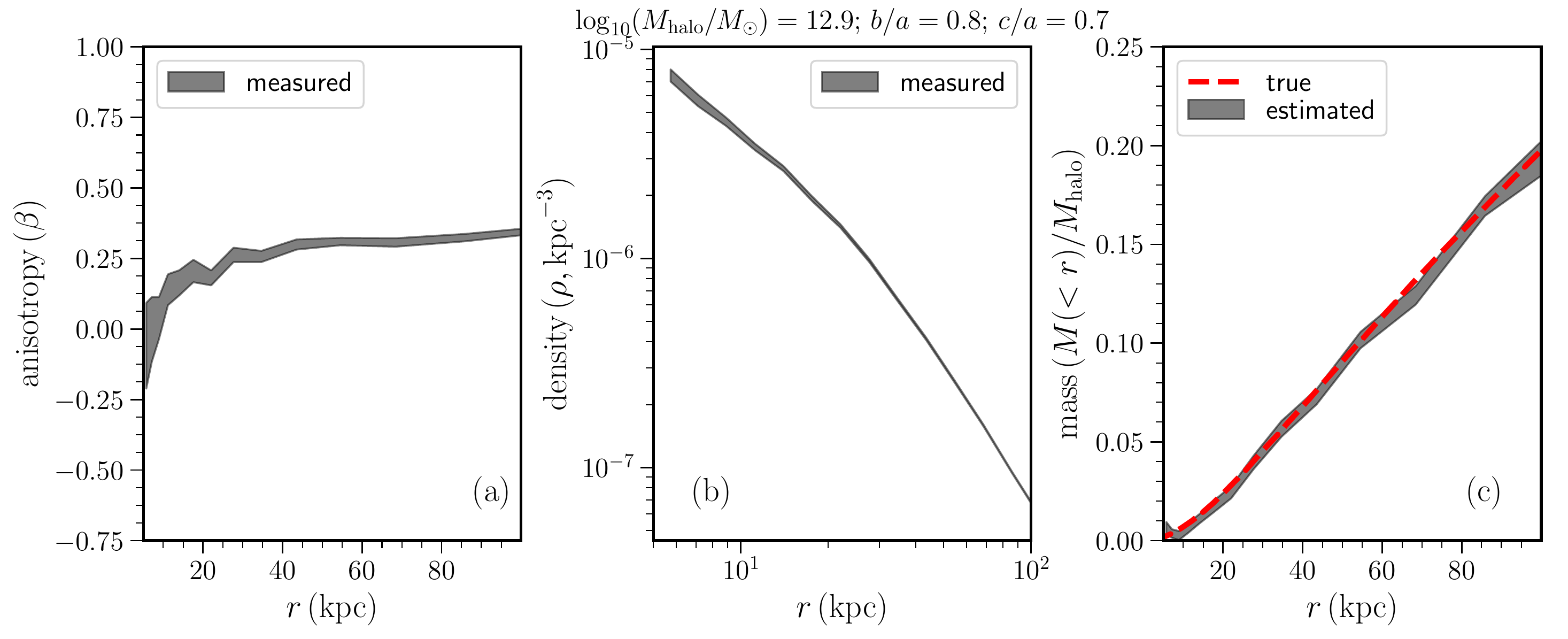}
    \caption{Spherical Jeans formalism in a triaxial DM halo from the \surfs\ simulation with velocity anisotropy, number density and cumulative mass profiles of the halo shown in panels a, b and c respectively. The grey bands show the measured quantities with bootstrapped uncertainties whereas red-dashed line in panel (c) represents the intrinsic mass distribution of the halo.}
    \label{fig:triaxialprofs}
\end{figure*}

\begin{figure*}
    \centering 
       \includegraphics[width=1\columnwidth]{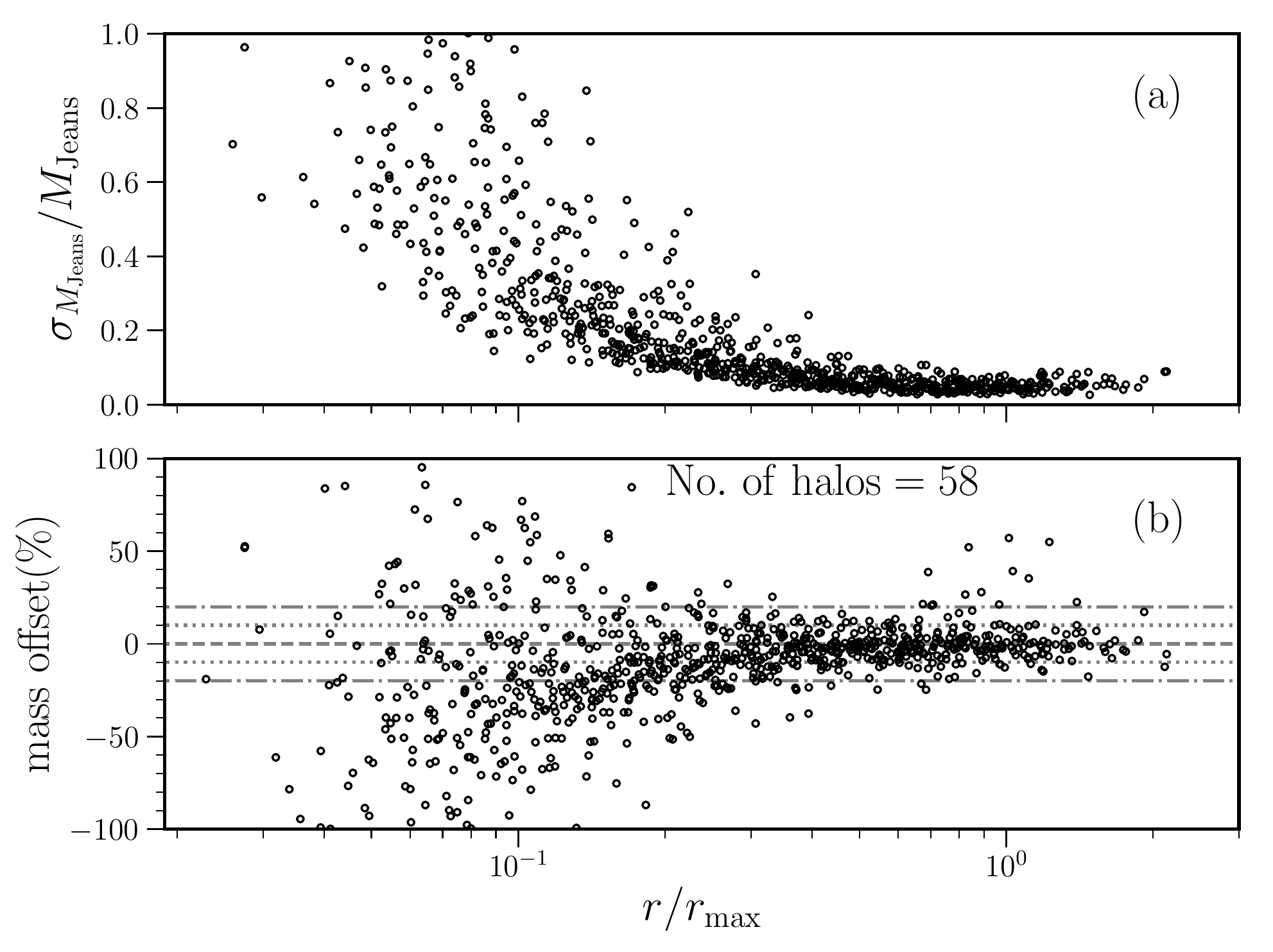}
       \includegraphics[width=1\columnwidth]{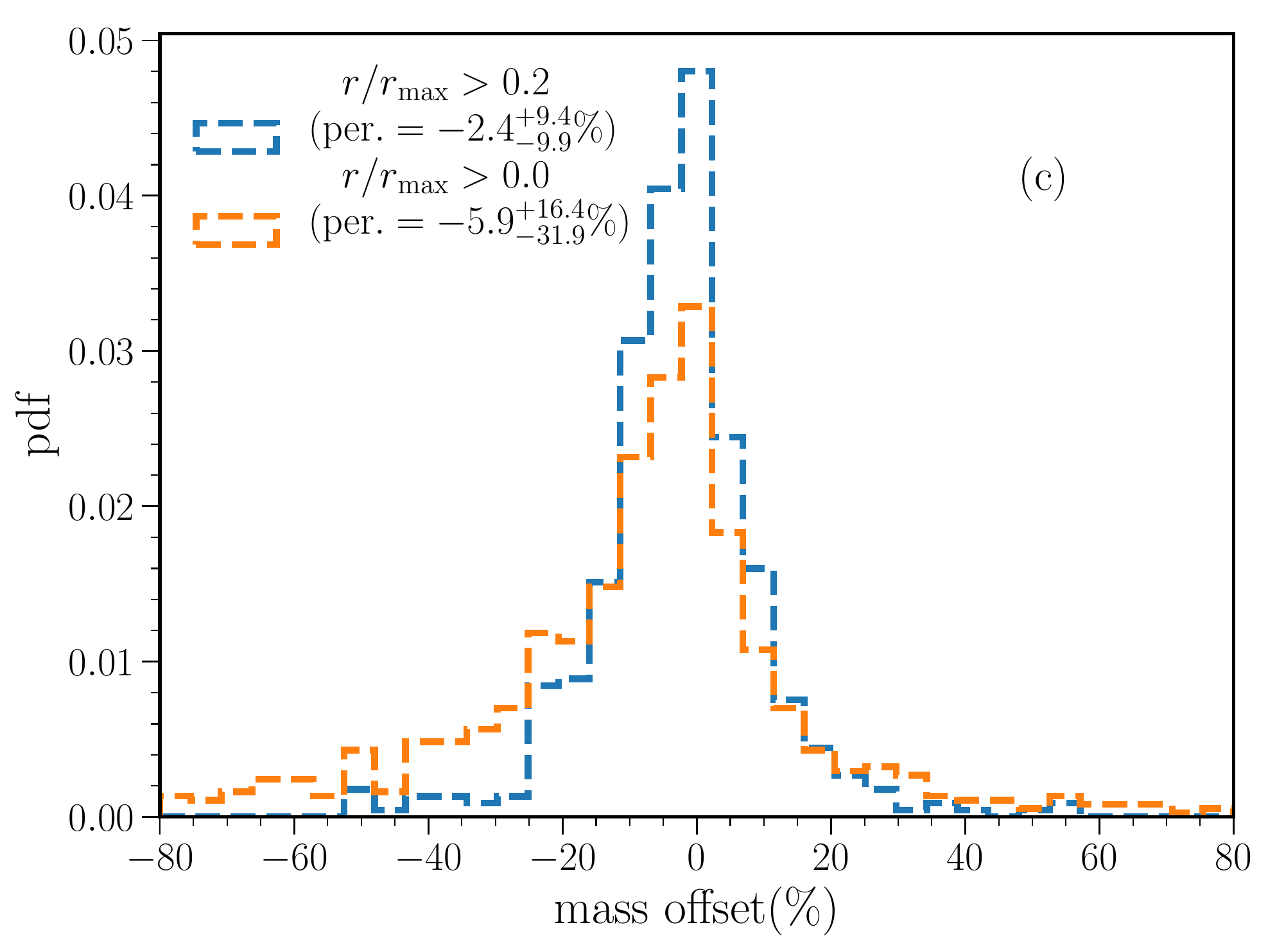}
       \includegraphics[width=2\columnwidth]{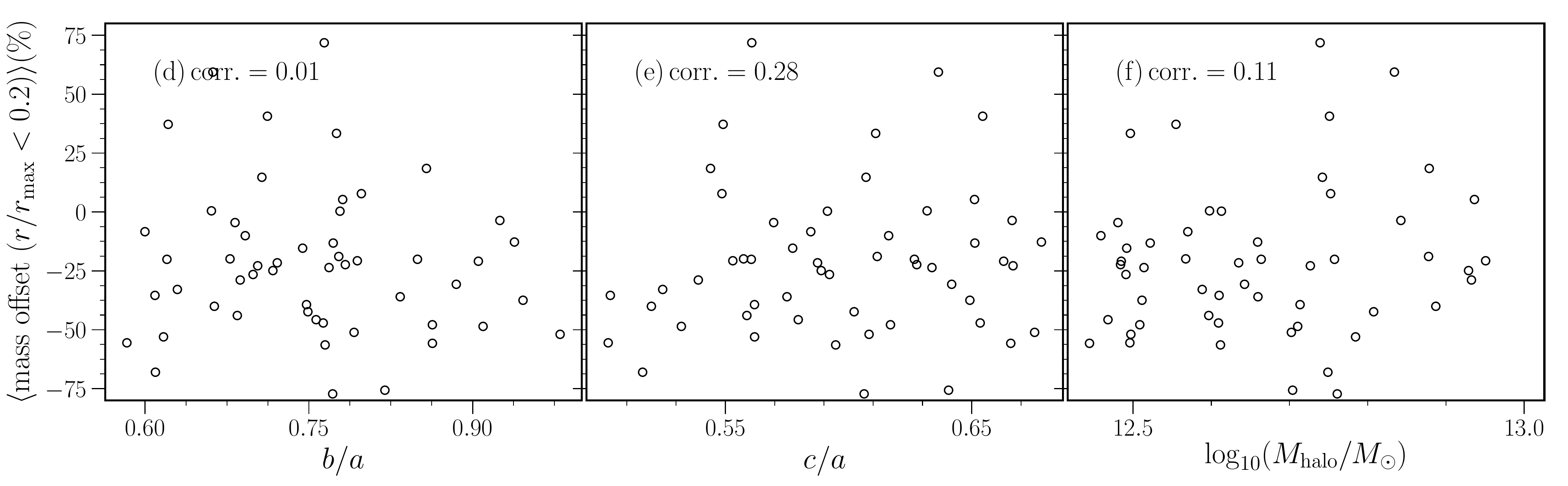}
    \caption{Error analysis of the mass measurement of triaxial DM halos taken from the \surfs\ simulation. Panel (a) shows the random uncertainties in mass measurements and panel (b) shows the percentage mass offset both as a function of scaled radius for all 58 halos whereas panel (c) shows the distribution of the mass offset. Panels d-f show the correlation of the mass offset with triaxiality parameters and the mass of the halo.}
    \label{fig:triaxialerrs}
 \end{figure*}
A generic prediction of structure formation under the \LCDM\ paradigm is that galactic DM halos are triaxial \citep[e.g.][]{2002ApJ...574..538J,2005ApJ...627..647B,2006MNRAS.367.1781A,2014MNRAS.439.2863V}, 
which has been shown to become comparatively more spherical due to influence of the baryonic processes 
\citep{2004ApJ...611L..73K,2006PhRvD..74l3522G,2010ApJ...720L..62K}.
The observations of the Galaxy show varying results. 
In this, study of debris of the Sagittarius dwarf tidal stream distributed on a great circle \citep{2001ApJ...551..294I} and also, the bifurcation in the stream \citep{2006ApJ...651..167F} 
suggest near spherical Galactic DM halo.
In contrast, the line-of-sight velocities of the stream favour a triaxial Galactic halo \citep{2004MNRAS.351..643H}.
More recently \cite{2009ApJ...703L..67L,2013MNRAS.428..912D}, while fitting spatio-kinematics of the Sagittarius stream find Galactic potential 
consistent to be triaxial and determine halo intermediate/major ($b/a$) axis ratio of 0.83 and minor/major ($c/a$) axis ratio of 0.67.
The constraints on the shape of the MW DM halo is still an unresolved subject; 
for additional constraints and further discussion of the topic see the review by \cite{2014JPhG...41f3101R}.

As discussed earlier the DM halo potential of the parent galaxy in \citetalias{2005ApJ...635..931B} simulations is spherical.
Therefore, to investigate the applicability of the spherical Jeans equation in the case of triaxial halos, here we utilise 
the \surfs\ simulation \citep{2017arXiv171201988E}.
The \surfs\ simulation is a suite of cosmological N-body simulations and we focus on its subset with box sizes 40$h^{-1}$ Mpc.
The halo catalogues are constructed with the VELOCIraptor phase-space halo finder \citep{2011MNRAS.418..320E}.
From the \surfs\ halo catalogue we focus on low mass groups with virial masses 
\footnote{mass enclosed within the radius where the over density is 200 times the critical density of the universe} of $\sim 10^{12.5}\,M_\odot$, 
which provides us with a sufficient statistical sample of well resolved halos composed of $\sim 5\times10^4$ DM particles.
Unfortunately, the \surfs\ simulation does not readily provide stellar tracer population and therefore, hindering a more realistic test.
As such it is not crucial at this point as we are only interested in understanding the sensitivity 
of the spherical Jeans equations in reconstructing the mass profiles of triaxial systems.
We note that the mass range of the selected halos we assume is slightly larger than a typical observational estimate of $\sim 10^{12}\,M_\odot$ for the virial 
mass of the Galaxy \citep[e.g.][]{2011MNRAS.414.2446M,2014ApJ...794...59K,2014A&A...562A..91P,2016ARA&A..54..529B},
a result of requiring haloes resolved with several tens of thousands of particles and the mass resolution of the simulation used.
This mass difference between our mock haloes and the MW is irrelevant as in this narrow mass interval we expect that the number density and kinematic profiles of the halos tracer populations remain self-similar.
Critically, here we are only testing the applicability of the spherical Jeans formalism in mass reconstruction of the non-spherical halos,
therefore all we need is a set of realistic triaxial halos.

Haloes within this mass range span a range of triaxiality parameters\footnote{the shape is calculated using the reduced inertia tensor \citep{1991ApJ...378..496D,2006MNRAS.367.1781A}},
from $b/a=[0.35,0.95]$ with typical values of 0.78 and with $c/a=0.7\,b/c$ \citep[][see Fig.7]{2017arXiv171201988E}.
In the end, we only use halos that have the triaxiality parameters reasonably close to the observed values of $b/a \approx 0.8$ and $c/a \approx 0.7$ 
for the MW DM halo \citep{2009ApJ...703L..67L,2013MNRAS.428..912D}.
The distributions of the axis ratio and halo mass of our final sample of 58 halos are shown in the top panels of the Fig.~\ref{fig:triaxialprops}.
Finally, we split the DM tracer population of these halos into 15 concentric spherical radial shells within $r\leqslant100\,\kpc$, 
and apply the spherical Jeans formalism (Section~\ref{sec:jeans}).
The grey bands in Fig.~\ref{fig:triaxialprofs} (a) and (b) show the measured anisotropy and number density of the tracer population whereas 
panel (c) shows the derived cumulative mass profile of the underlying DM halo with the total mass of $\log_{10}(M_\text{halo}/M_\odot) = 12.9$ and axial ratios of $b/a=0.8$ and $c/a=0.7$.
In panel (c) the over-plotted red dashed-line represents a spherically averaged intrinsic mass profile of the halo.
We see that the spherical Jeans formalism reconstruct the mass profile of the halo reasonably well and results of the remaining 57 halos are also consistent with this.

In Figure~\ref{fig:triaxialerrs} we provide an error analysis of the estimated masses of the triaxial halos.
The panel (b) demonstrates the mass offset (in percentage) for all the 58 halos as a function of scaled radius $r/r_\text{max}$, where $r_\text{max}$ corresponds to the radius $r$ at which the circular velocity of the halo is maximum.
Dashed lines in the panel highlights the offset of $10\%$ and $20\%$.
The blue and orange histograms shown in panel (c) however show the distributions of mass offset at all scaled-radii $r/r_\text{max}>0.2$ and $>0.0$ respectively.
The median mass offset at $r/r_\text{max}>0$ is $-5.9^{+16.4}_{-31.9}\%$, and $-2.4^{+9.4}_{-9.9}\%$ at $r/r_\text{max}>0.2$.
The dispersion in the mass offset in the case of DM tracer populations of the 11 \LCDM\ halos of \citetalias{2005ApJ...635..931B} at outer region ($r\geq10~\kpc$) is of similar level $\sim 7 \%$ although the bias is positive here ($\sim 3\%$).
The dispersion obtained for $r/r_\text{max}>0.2$ case is $8.7\%$ after adjusting for random uncertainty.
In agreement with the \citetalias{2005ApJ...635..931B} \LCDM\ halos, here again we confirm that the biases in the mass measurements are not due to subtracting mean motion while calculating the velocity dispersion (in Equation~\ref{eqn:veldisp}). 
In the figure we observe that generally in the inner $r/r_\text{max}<0.2$ region the scatter in the residual of the mass estimate is large.
To further investigate the anomaly, in panels (d), (e) and (f) we show the relation of average mass offset at $r/r_\text{max}<0.2$ as a function of $b/a$, $c/a$ and halo mass respectively, and also provide their respective correlation measurements.
We fail to find any correlations between the observed mass offsets and the aforementioned intrinsic properties of the halos.
Finally, from panel (a), where the ratios of dispersion to the estimated mass $M_{\mathrm{Jeans}}$ of the triaxial halos are shown, and similar to the earlier cases of \citetalias{2005ApJ...635..931B} \LCDM\ halos, here also we find that the large scatter in the residual of the mass estimate in the inner region is mainly due to Poisson noise. 

In summary, we conclude that in the outer $r/r_{\mathrm{max}}>0.2$ the underlying mass profiles of the \surfs\ halos can be determined with a bias of $\sim-2.4\%$ and a dispersion of $\sim10\%$. Note, this dispersion includes the effects of triaxiality as well as departures from equilibrium.

\section{Discussion and conclusion}\label{sec:conclusions}
In this paper, we utilise the 11 Milky Way stellar halos simulated in accordance with \LCDM\ by \citetalias{2005ApJ...635..931B}, 
6 additional simulated stellar halos from \citetalias{2008ApJ...689..936J} built to have artificial accretion histories dominated by events that are predominantly recent/old, 
on radial/circular orbits or having larger/smaller satellite mass, and 58 triaxial DM halos obtained from the \surfs\ simulation \citep{2017arXiv171201988E} 
to test the efficacy of the spherical \cite{1915MNRAS..76...70J} formalism in predicting the mass distribution of the Milky Way analogues in a \LCDM\ universe.

In overall, using the spatio-kinematic profiles of the stellar tracer population and the spherical Jeans equation we recover the underlying mass distribution of the Milky Way analogues in a \LCDM\ universe within $r/\kpc\in[10,100]$, with a bias of $\sim 12\%$ and a dispersion of $\sim14\%$ ($12\%$ when adjusted for random uncertainty).
Additionally, analysing triaxial DM halos obtained from the \surfs\ simulation with intermediate/major axis ratio in range [0.5, 1] and minor/major axis ratio in range [0.5, 0.7], we are able to recover the underlying mass distribution of the halos with a bias of $\sim -2.4\%$ and a dispersion of $\sim10\%$ ($8.7\%$ when adjusted for random uncertainty), in the outer $r/r_\text{max}>0.2$ region of the halos. 
Similar level of dispersion ($\sim7\%$ or $\sim6\%$ when adjusted for random uncertainty) is also observed for the case of the DM halos of  \citetalias{2005ApJ...635..931B} but with positive bias of $\sim 3\%$.
 
In perfect conditions, a spherical system in equilibrium, we can correctly reconstruct the mass profile and there is no bias and dispersion that can be accounted by random uncertainty on the estimated mass. When applied to data from simulations we do see some bias and dispersion, meaning the bias and dispersion can be either due to system being aspherical and out-of-equilibrium or due to an unknown effect in simulations. We find little correlation of mass offsets with asphericity, so out-of-equilibrium effects seem to be the main cause behind the observed mass offsets. The out-of-equilibrium effects will shift the mass estimates at any given radius and this will lead to a non-zero dispersion. So the dispersion in principle sets a limit on the accuracy with which we can expect to measure mass at a given radius for any halo using the spherical Jeans equation. This limit is set due to the inherent nature of the \LCDM\ halos and is independent of the quality and the quantity of the observational data.

Using {\sc galaxia} \citep{2011ApJ...730....3S}, a stellar population synthesis software utilising Padova isochrones \citep{2008A&A...482..883M}, we estimate that, till the magnitude limit of 17 in V-band, {\it Gaia} will have more than 5 times the number of tracers that we have investigated here. So, the error on the mass estimates of the Galaxy using stellar tracers provided by the {\it Gaia} using the Jeans analysis will be limited by the non-equilibrium effects as well as uncertainties in observed distances and tangential velocities rather than the sample size.

The fact that the two different simulations (with DM particles as tracers) give similar level of dispersion ($10\%$ for \surfs\ and $7\%$ for \citetalias{2005ApJ...635..931B} halos) in estimated masses is reassuring for our estimate of the dispersion.  When applied to \citetalias{2005ApJ...635..931B} stellar halos (stellar particles as tracers) we get a higher dispersion ($14\%$). This is the case that is of practical use, as we observe stars rather than dark matter particles. The higher dispersion here is also as expected, since, the stellar halos have more substructures than dark matter halos. Dispersion could also be due to random uncertainty associated with our mass estimator (effect of Poisson noise).  We have shown that the random uncertainty is a factor of two smaller than the measured dispersion for all analysed cases. The total variance being the sum of squares of random and intrinsic scatter, the random uncertainty should make very little contribution to the dispersion. Taking the random uncertainty into account we estimate the intrinsic scatter due to non-equilibrium effects to be $12\%$ for the case of stellar halos.

The interpretation of bias is less obvious. Naively we expect the bias to be zero, at a given radius, the non equilibrium effect can shift the mass in either direction (as the quantities on the right side of equation~\ref{eqn:jeanseqn} can shift in either direction). 
For the \surfs\ as well as \citetalias{2005ApJ...635..931B} dark matter halos, although not negligible but the bias is a factor of two smaller than the dispersion. For the \citetalias{2005ApJ...635..931B} stellar halos the bias is much higher but still less than the dispersion.  
The sign of the bias is different for the \surfs\ and \citetalias{2005ApJ...635..931B} simulations, and we have not been able to find a reason for it. 
However, the bias is significant only if we treat the measurement at each radius as independent and this might not be true. 
Typically, a few luminous/massive accretion events dominate in shaping the overall properties of a given the stellar halo.  
Specifically the fraction of material in substructures is dominated by a few luminous accretion events \citep[Sec 4.4]{2011ApJ...728..106S}. So mass measurements of a given halo at different radius can be correlated. Given that we only have 58 independent halos for the \surfs\ and 11 for \citetalias{2005ApJ...635..931B}, the measured biases are less than twice of (disperion/$\sqrt \textrm{number of halos}$) and are within the $2\sigma$  limit. For the case of stellar halos the bias is higher but still within $2.5\sigma$ limit. We note that in \citetalias{2005ApJ...635..931B} simulations, a stellar halo is created by assigning unequal weights to dark matter particles. 
Although less likely (as after getting disrupted the particles of a satellite behave more or less independently) but it is worth exploring in future if this weighting scheme can introduce a systematic bias. Another possibility could be that the bias is a reflection of the initial conditions  used to generate the accretion history of the halo. 
The accreting satellites tend to be closer to (or further from) peri- or apo-centre than one would expect for a phase mixed population, meaning that their radial velocities are lower (or higher) than in that case.
When the satellites disrupt and get fully phase mixed the system will be in equilibrium and the bias would vanish. However, here we are analysing a partially relaxed system, as evidenced by the presence of  significant amount of substructure, and this can  can lead to a non zero bias.  
For example, if we consider the simple spherical collapse model of the formation of a dark matter halo, at turn around the total energy is mostly potential, while at the collapse it is kinetic. 
Applying Jeans equation here will give systematically lower mass at the turn around stage and higher mass at the collapse stage.  
The fact that the initial conditions are different in \surfs\ and \citetalias{2005ApJ...635..931B} simulations, could possibly lead to different biases in them.

Additionally, the investigation of the simulated stellar halos with artificial accretion history suggests that the young halo built from events less than 8 Gyr and the halo dominated by high luminosity ($>10^7$ L$_\odot$) accretion events, show the most undulations in the mass profiles and hence, are the most error prone cases to apply the Jeans analysis. This is due to significant amount of unrelaxed substructures inherently present in these halos. On the contrary, a halo dominated by less luminous ($<10^7$ L$_\odot$) accretion events, a well mixed halo, provide good scenario to apply the Jeans formalism. 

\section*{ACKNOWLEDGEMENTS}
PRK is funded through Australian Research Council (ARC) grant DP140100395 
and the University of Western Australia Research Collaboration Awards PG12104401 and PG12105203. We like to thank the referee for constructive and insightful comments that helped to improve the paper significantly. We also thank Prof. Chris Power and Dr. Claudia Lagos for providing the \surfs\ data, and Prof. Geraint Lewis and Dr. Luke Davies for discussions related to the paper.

\emph{Software credit}: 
{\sc ipython} \citep{ipython}, {\sc matplotlib} \citep{matplotlib}, 
{\sc seaborn} \citep{seaborn}, 
{\sc pandas} \citep{pandas}, {\sc numpy} \citep{numpy} and {\sc scipy} \citep{scipy}.

\section{Appendix}\label{sec:dmcase}
\subsection*{The case of accreted DM tracer population}
\begin{figure*}
  \centering
 \includegraphics[width=0.9\columnwidth]{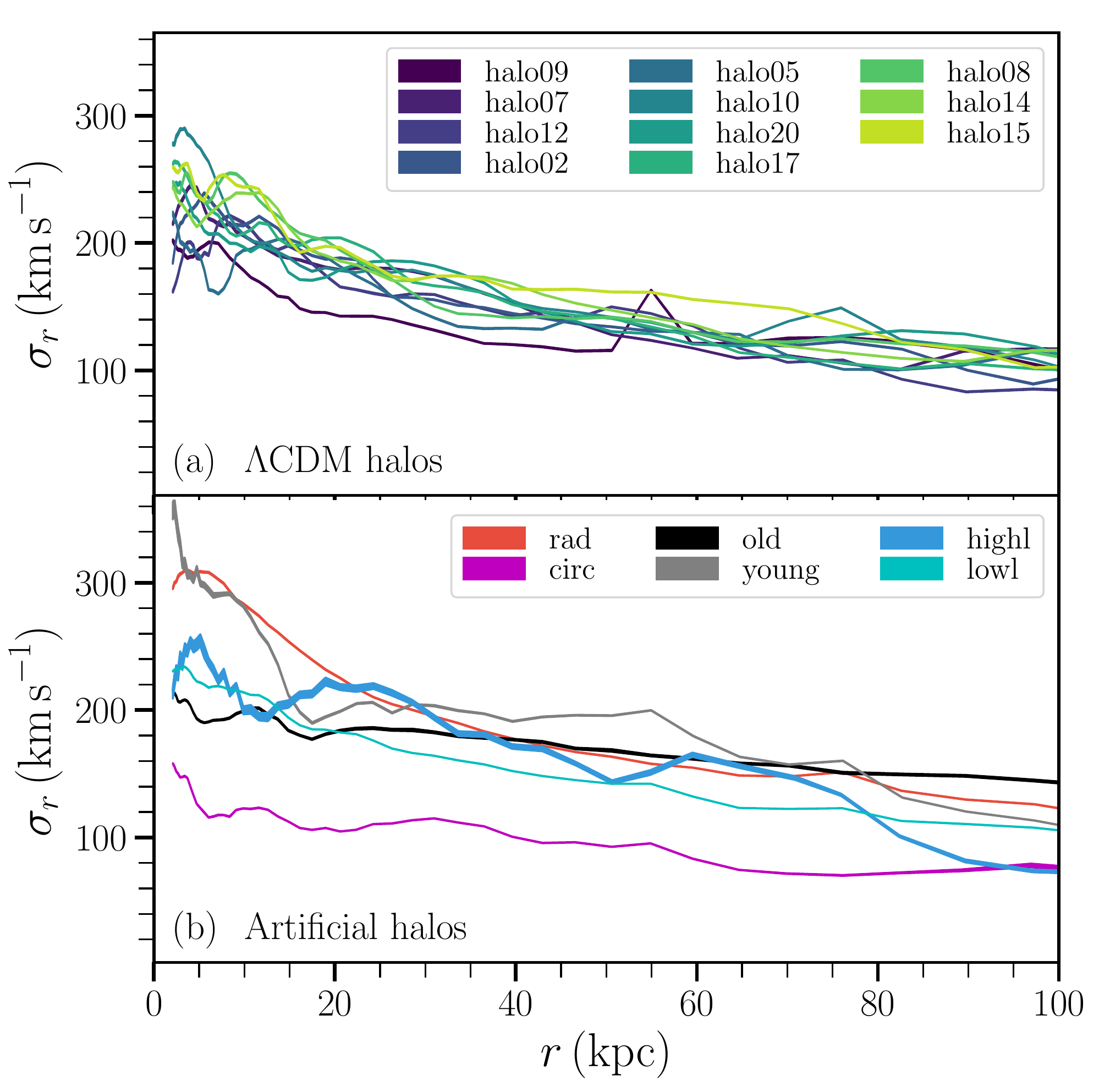}
 \includegraphics[width=0.9\columnwidth]{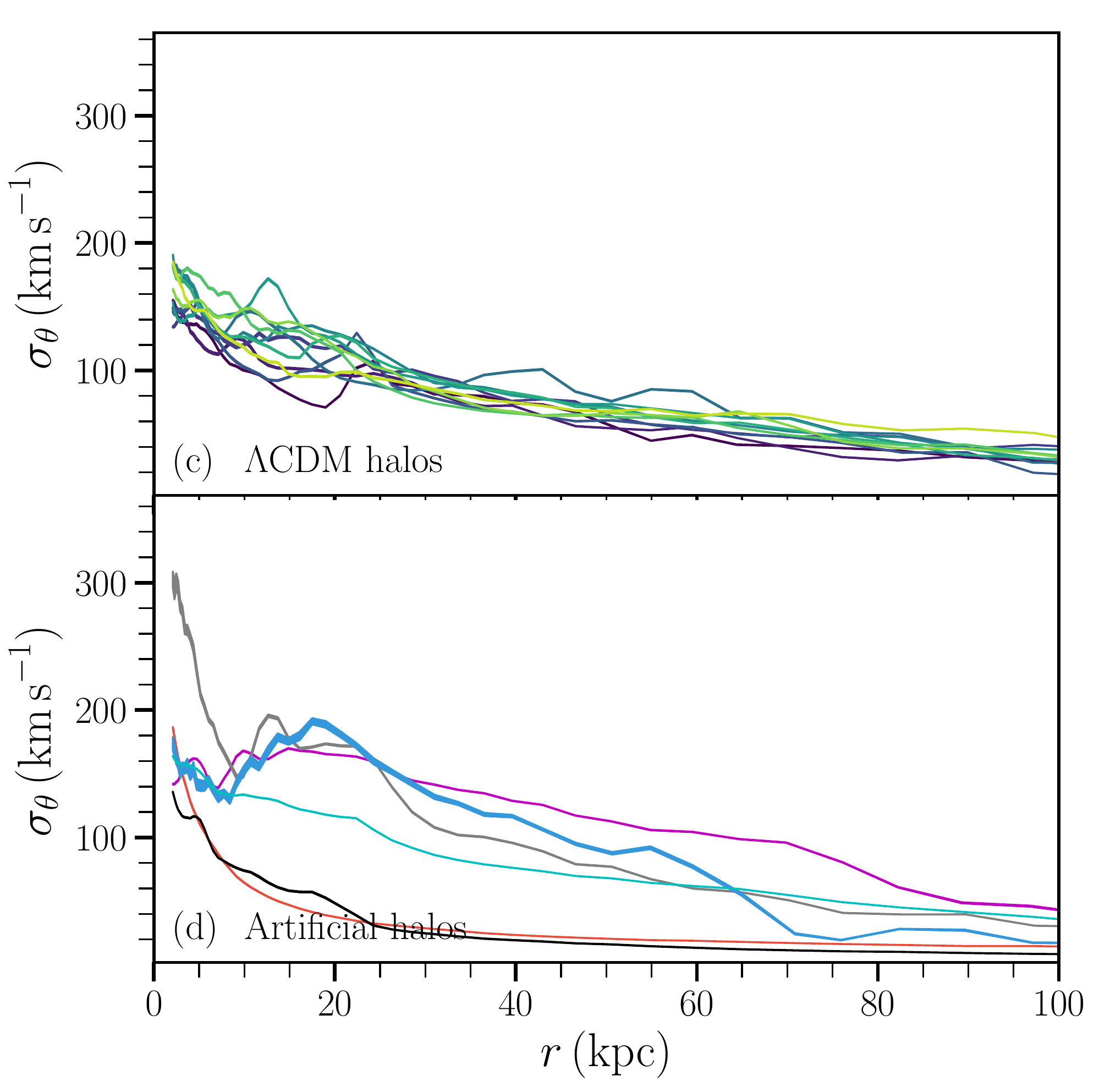}
 \includegraphics[width=0.9\columnwidth]{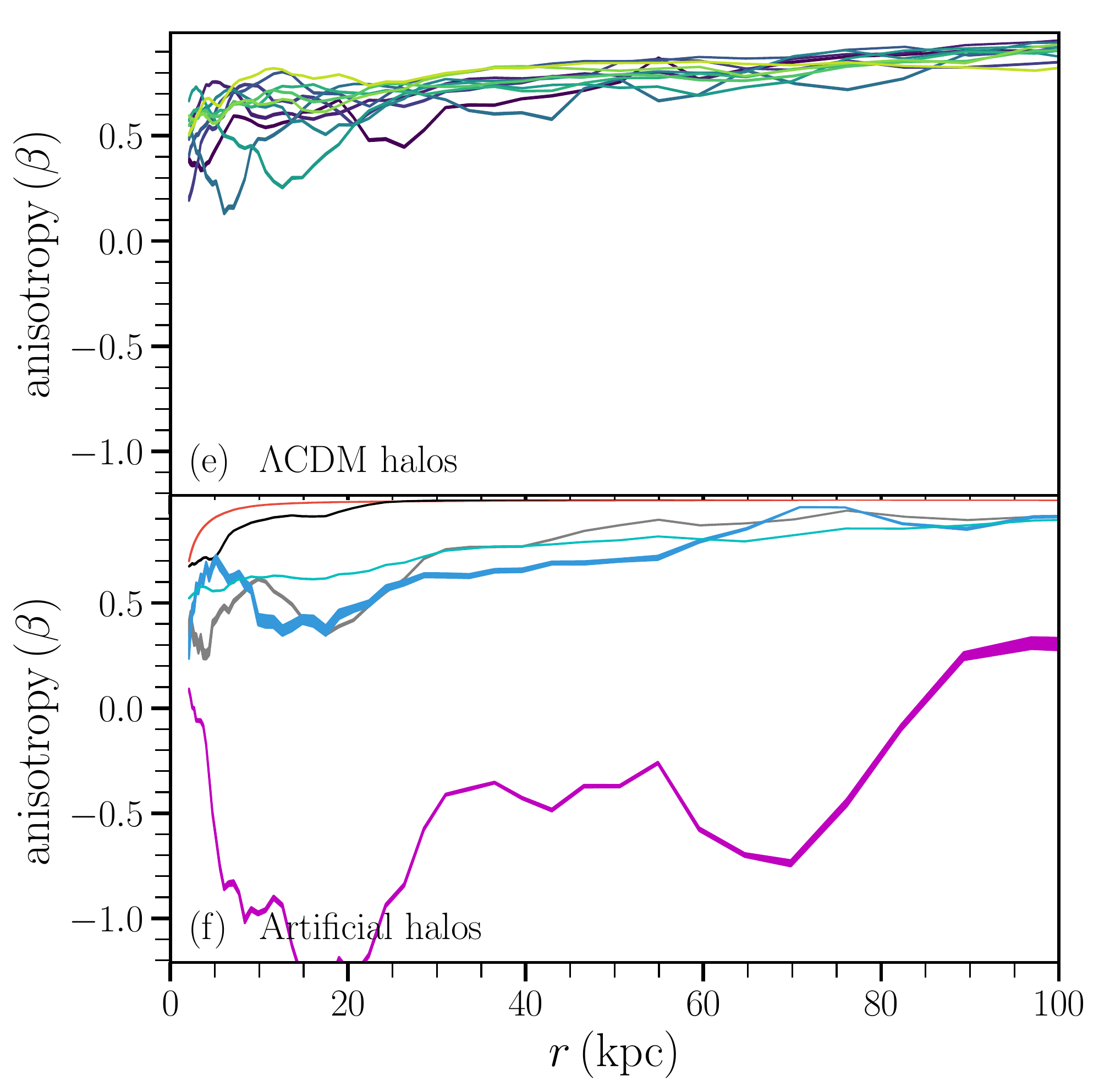}
 \includegraphics[width=0.9\columnwidth]{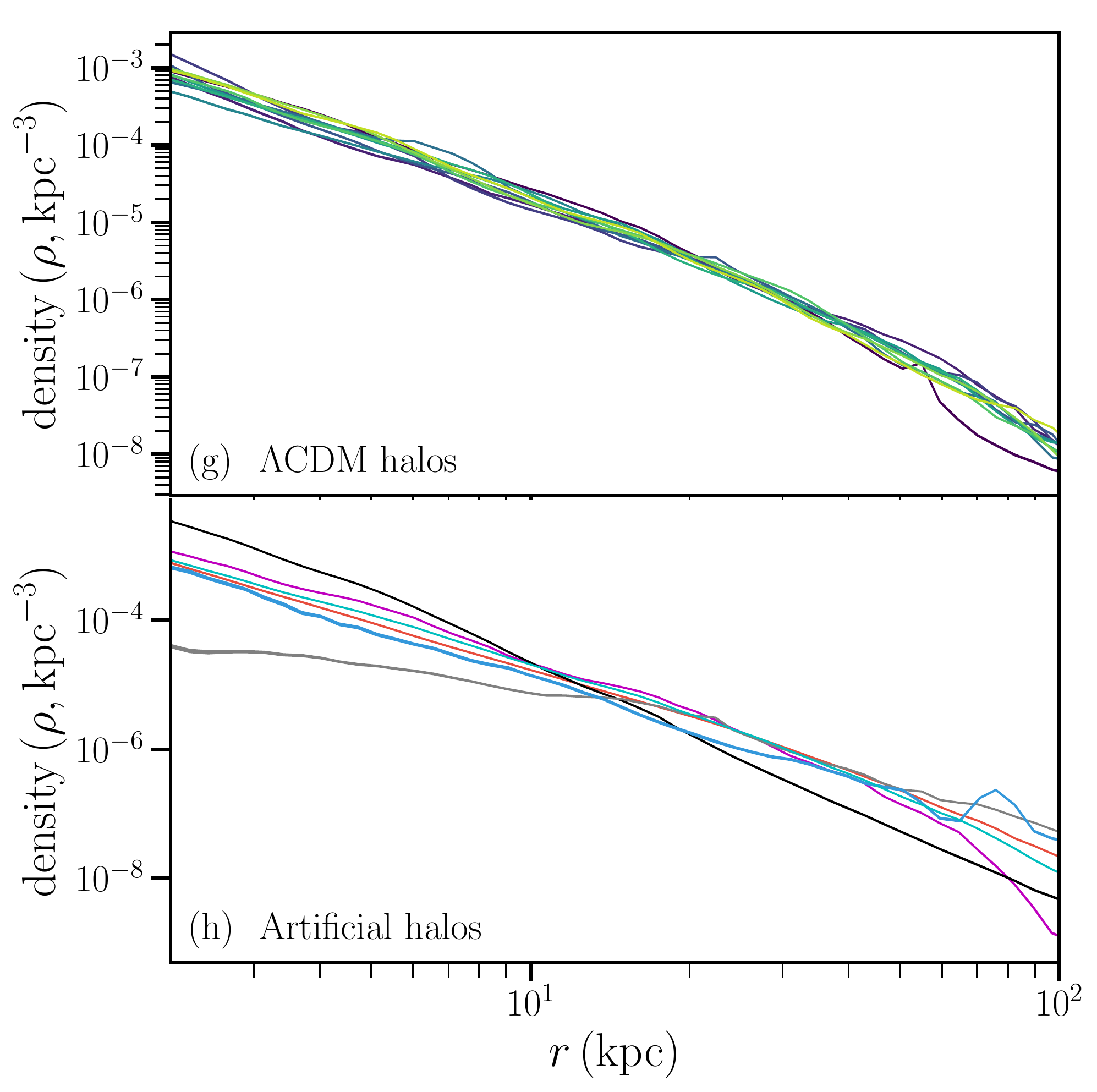}  
 \includegraphics[width=0.9\columnwidth]{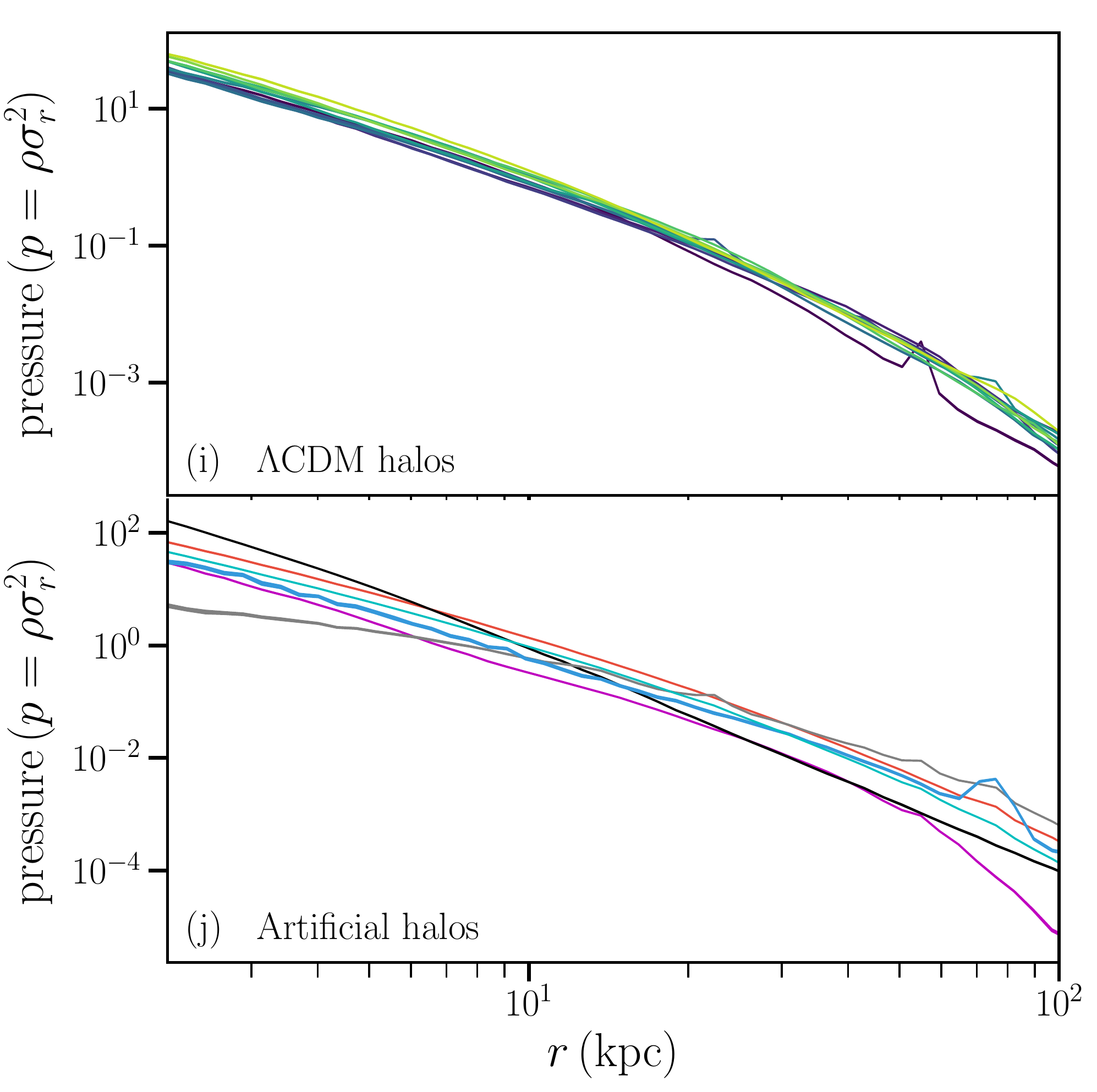}
 \includegraphics[width=0.9\columnwidth]{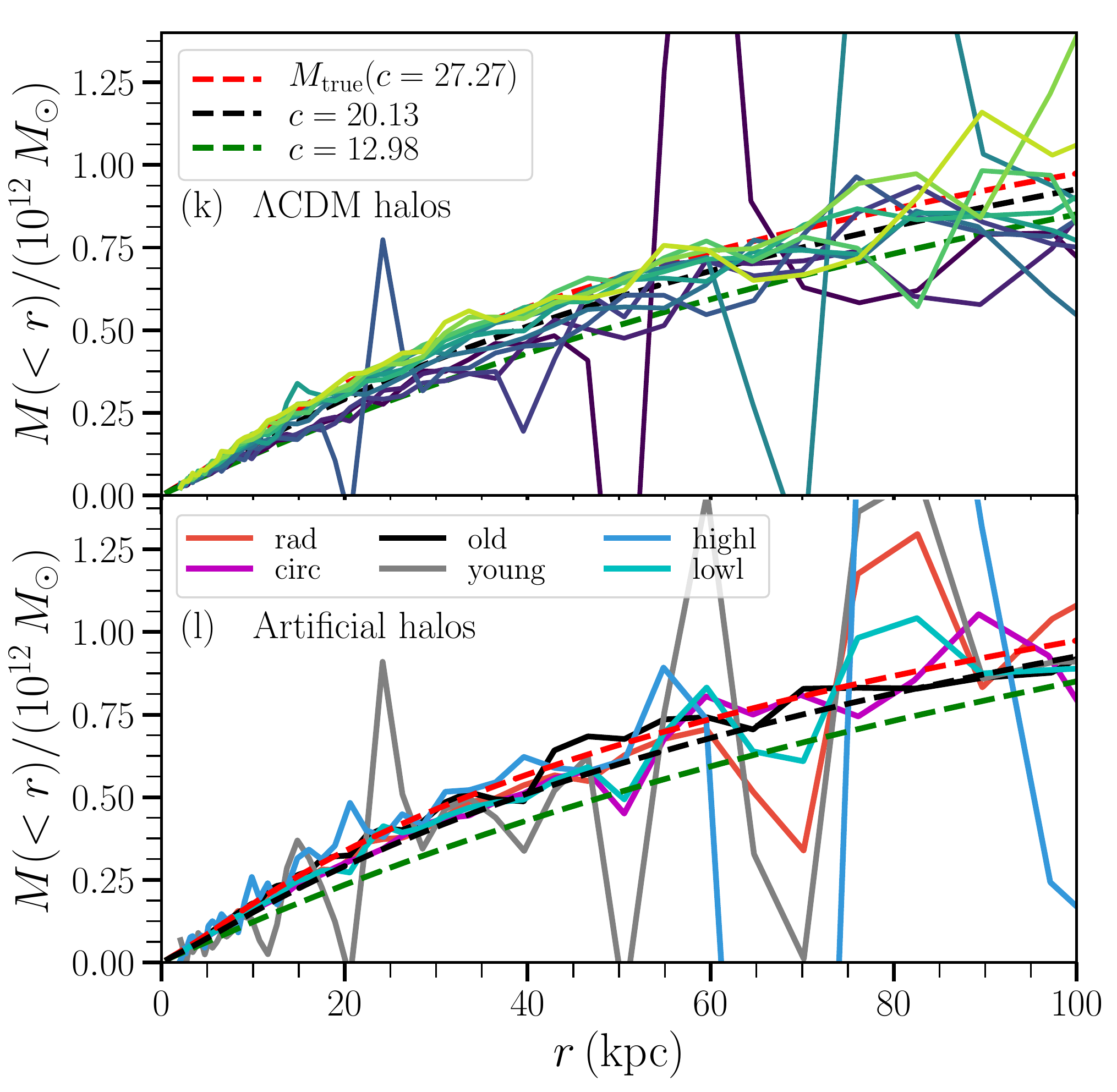}
  \caption{Analogous to Fig.~\ref{fig:profiles_wt}, but for the case when accreted DM particles taken from \protect\citetalias{2005ApJ...635..931B} and \protect\citetalias{2008ApJ...689..936J} simulations are used as a dynamical tracer.}
  \label{fig:profiles}
\end{figure*}
\begin{figure*}
  \centering 
  \includegraphics[width=1\columnwidth]{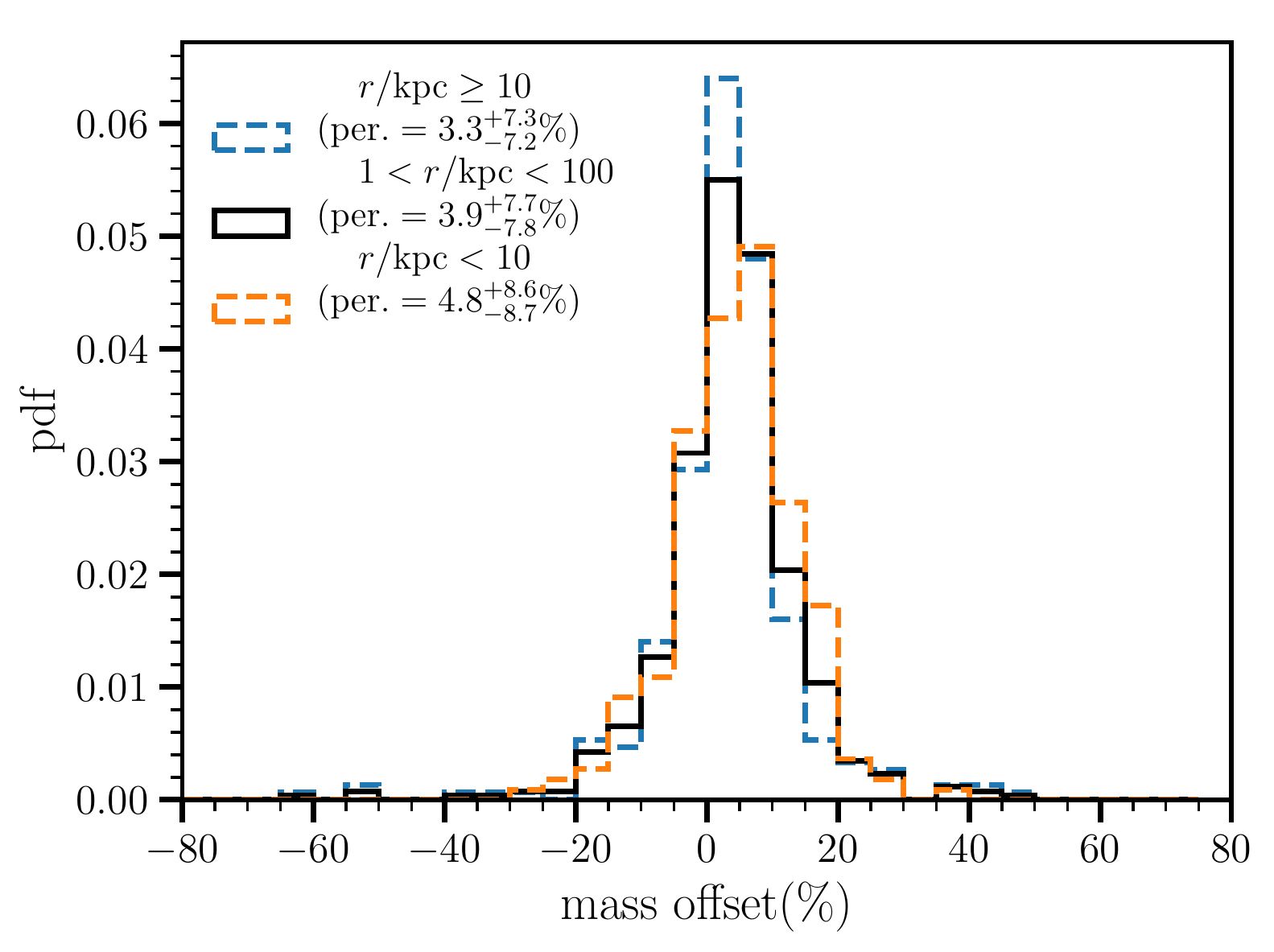}
    \caption{Error analysis of the estimated masses of the parent galaxy using the accreted DM tracer populations of the 11 \LCDM\ halos,
  with labels same as Fig.~\ref{fig:lcdm_mpe}(c).}
  \label{fig:lcdm_dm_mpe}
\end{figure*}
Here we demonstrate key results obtained using the accreted DM tracer population although this may not have any observational relevance as it is not possible to directly observe the DM. 
Nonetheless, we conduct the exercise to establish a few point discussed below.
First, in Figure~\ref{fig:profiles} (analogous to Fig.~\ref{fig:profiles_wt}) we demonstrate key spatio-kinematic properties of 
the accreted DM tracer population (with sample size of $10^5$) from \citetalias{2005ApJ...635..931B} and \citetalias{2008ApJ...689..936J} simulations.
We note that similar to Fig.~\ref{fig:profiles_wt}, where stellar tracer populations are utilised, both the $\rsigma$ and $\thsigma$ (also $\phsigma$, 
not shown in the figure) attain highest value for small $r$ and vice-versa, which turn-over at $r \simeq 5$ kpc.
Similarly, except the case of {\it circ} halo all the simulated stellar halo have predominantly radial orbits.
However, we note that compare to the earlier case when stellar tracers are used, the velocity dispersion profiles and hence, also anisotropy, have larger inter-halo scatter.
Notably, we also find that the density slope at $r<$ break-radius is comparatively steeper in the current case.
Second, the corresponding reconstructed mass profiles of the individual halos are already shown with blue lines in Fig.~\ref{fig:massprofile}.
Finally, in Fig.~\ref{fig:lcdm_dm_mpe} (analogous to Fig.~\ref{fig:lcdm_mpe} c) we present the distributions of the biases in the estimated masses at various distance range.

Number of DM tracers we utilise in Fig.~\ref{fig:massprofile} are roughly of the similar order of magnitude to that of the stellar tracers we had in Section~\ref{sec:results}, i.e., $10^5$. 
We find that in this case the overall mass offset is $\sim3\%$ with $\sim7\%$ ($\sim6\%$ when corrected for random uncertainty) of dispersion, 
which are clearly better than the respective errors of $12\%$ and $14\%$ we obtained earlier with the stellar tracer populations.
This is expected and has a physical cause. 
As mentioned in Section~\ref{sec:data}, in the \citetalias{2005ApJ...635..931B} simulation the stars are ``painted on'' by assigning a luminosity weight to each dark matter particle within an accreted satellite.
These luminosity weights are proportional to the binding energy of the satellite meaning the stellar tracer populations generally comes from the core of the accreted satellites, 
hence, takes longer to relax compare to the DM tracer population resulting larger upset in the estimated mass profiles.

\bibliographystyle{mnras}
\bibliography{all}

\bsp	
\label{lastpage}
\end{document}